\documentclass[review, 5p]{elsarticle}

\usepackage{mathptmx}
\usepackage{amsmath}
\usepackage{amssymb}
\usepackage{graphicx}
\graphicspath{{figures/}}
\usepackage{float}
\usepackage{dsfont}
\usepackage{amsfonts}
\usepackage[T1]{fontenc}

\usepackage{xcolor}
\usepackage{setspace}

\usepackage{url}
\usepackage{stfloats}
\usepackage{subfig}

\usepackage{stmaryrd}

\usepackage{algorithm}
\usepackage{algpseudocode}

\usepackage[normalem]{ulem}
\newcommand{\add}[1]{\textcolor{black}{#1}}
\newcommand{\del}[1]{}

\usepackage{lineno}

\usepackage[colorinlistoftodos]{todonotes}
\usepackage{listings}

\usepackage{hyperref}

\author[a,b]{Alexandre Blain}
\address[a]{Inria, CEA, Universit\'e Paris-Saclay,
            Paris, France}

\author[a]{Bertrand Thirion}
\author[b]{Pierre Neuvial}

\address[b]{Institut de Math\'ematiques de Toulouse, UMR 5219, Universit\'e de Toulouse, CNRS,
UPS, Toulouse, France}

\begin{document}

\journal{NeuroImage}

\begin{frontmatter}
\title{Notip: Non-parametric True Discovery Proportion control for brain imaging}
\begin{abstract}
Cluster-level inference procedures are widely used for brain mapping. 
These methods compare the size of clusters obtained by thresholding brain maps to an upper bound under the global null hypothesis, computed using Random Field Theory or permutations. 
However, the guarantees obtained by this type of inference - i.e. at least one voxel is truly activated in the cluster - are not informative with regards to the strength of the signal therein. 
There is thus a need for methods to assess the amount of signal within clusters; yet such methods have to take into account that clusters are defined based on the data, which creates circularity in the inference scheme.
This has motivated the use of \emph{post hoc} estimates that allow statistically valid estimation of the proportion of activated voxels in clusters. 
In the context of fMRI data, the All-Resolutions Inference framework introduced in \cite{rosenblatt2018all} provides post hoc estimates of the proportion of activated voxels. 
However, this method relies on parametric threshold families, which results in conservative inference. 
In this paper, we leverage randomization methods to adapt to data characteristics and obtain tighter false discovery control.
We obtain \emph{Notip}, \add{for Non-parametric True Discovery Proportion control}: a powerful, non-parametric method that yields statistically valid \del{estimation} \add{guarantees} \del{of} \add{on} the proportion of activated voxels in data-derived clusters.
Numerical experiments demonstrate substantial \del{power} gains \add{in number of detections} compared with state-of-the-art methods on 36 fMRI datasets. 
The conditions under which the proposed method brings benefits are also discussed. 

\end{abstract}
\end{frontmatter}

\section{Introduction}

The mapping of the human brain consists \del{in} \add{of} associating regions of the brain with cognitive functions or disorders. 
This is important both for basic neuroscience, e.g. the understanding of brain function, and medical applications, as it allows to identify regions that carry disease-related signal.
The most popular modality to map brain function is functional Magnetic Resonance Imaging (fMRI), as it is non-invasive and offers decent spatial resolution (about $2mm$ isotropic) and full brain coverage.\\

FMRI data are sampled on a discrete 3D lattice and subject to various preprocessing steps \cite{esteban2019}, resulting in a set of \emph{voxels} that contain a signal that reflects brain activity. 
After suitable statistical analysis, relevant brain territories can be reported.
More precisely, practitioners define a \emph{contrast}, that is, a linear combination of a set of images, typically corresponding to the comparison between two
or more conditions or groups of participants, and seek to test
hypotheses $\mathbf{H_{0,i}}$: "Voxel $i$ is \emph{inactive} for
this contrast", meaning that it does not show any effect for the
selected contrast, versus $\mathbf{H_{1,i}}$: "Voxel $i$ is
\emph{active} for this contrast".
This statistical problem entails a dire multiple testing issue as described in \cite{friston1991comparing}, as standard fMRI images comprise between $50k$ and $400k$ voxels (growing to millions with the development of high-resolution imaging).\\ 
%

In this context, if multiplicity is not accounted for, the number of false discoveries is unacceptably high.
%
%
In other words, mere voxel-wise type 1 error control is not appropriate in the context of multiplicity. Family-Wise Error Rate (FWER) control can be used in this setting \cite{friston1991comparing} but it is conservative, resulting in false negatives, which hurts reproducibility (see e.g. \cite{Thirion2007,Button2013}).
A more powerful and commonly used approach is to control the False Discovery Rate (FDR) \cite{genovese2002thresholding},
which is systematically done using Benjamini-Hochberg procedure \cite{benjamini1995controlling}. 
A caveat to this approach is that the FDR actually corresponds to the \textit{expected} False Discovery Proportion (FDP). 
The FDP is the proportion of false discoveries among all discoveries.
As noted by several authors \cite{genovese2004stochastic,korn2004controlling,neuvial2008asymptotic}, FDR control does not guarantee FDP control.\\

An alternative type of inference to increase statistical power is to perform inference at cluster-level, rather than voxel-level
\cite{poline1993analysis}, because brain activation is organised in compact regions \emph{(clusters)} in the brain volume. 
This type of inference tests whether regions above a given threshold are larger than expected under the null hypothesis, or whether the total amount of signal in these regions \cite{smith2009threshold} exceeds its expected value under a null distribution.
However, this approach suffers from several problems \cite{eklund2016cluster}, such as the arbitrary choice of cluster-forming threshold \cite{Woo2014}, or the difficulty to establish a null distribution for cluster size and aggregated signal.
To address this last issue, reliable non-parametric solutions have been proposed \cite{winkler2014,eklund2016cluster}. 
However, the arbitrariness regarding cluster-forming threshold is hard to deal with. 
To overcome it, one may define such clusters or regions, and \textit{then} assess the proportion of active voxels in each region, i.e the True Discovery Proportion, TDP = 1 - FDP.
Such a region of interest could be defined a priori, using an anatomical atlas, or a posteriori, based on the fMRI data. 
For instance, one might wonder what is the proportion of active regions in a \emph{blob}, i.e. a contiguous set of statistical values that are higher than the image background. 
Yet, such a definition of the clusters after seeing the data raises a double-dipping issue, which can lead to massive false positive inflation~\cite{kriegeskorte2009circular}.\\

To illustrate this statistical bias, let us consider a classical example of invalid post-selection inference. 
Users often perform a first round of tests to identify potentially interesting regions (i.e., regions comprising significant signal).
If inference is performed only on smallest $p$-values obtained at this first round, then the FDP is not controlled, as shown in \cite{blanchard2020post}.
To bypass this double-dipping issue, one can use \emph{post hoc} estimates that control the FDP. \add{Note that an upper bound on the FDP is equivalent to a lower bound on the TDP.}
The first method of that kind is a parametric method called All-resolutions inference (ARI) \cite{rosenblatt2018all}.\\

In this paper, we introduce the Notip procedure, that adapts non-parametrically to data correlation \del{structure}. 
\add{The use of non-parametric procedures also renders the inference robust to mis-specification of the statistics distribution.}
We study whether such a procedure can yield less conservative inference while offering the same statistical guarantees.
We perform extensive experiments on dozens of fMRI datasets to compare the number of detections obtained by this approach with that of existing methods.\\

The paper is organized as follows. 
In Section \ref{sec:fdp-control-by-jer}, existing methods for the post hoc control of FDP are introduced via the notion of \emph{Joint Error Rate} (JER) proposed by \cite{blanchard2020post}.
Our main contribution is the Notip method presented in Section \ref{sec:data-driven-templates}: a nonparametric data-driven approach that relies on the JER framework to obtain sharper post hoc FDP control.
Numerical experiments and results on fMRI data reported in Sections \ref{sec:experiments} and \ref{sec:results} show that substantial \del{power} gains \add{in the number of detections} are obtained from the proposed method, while controlling the FDP of the detected regions at a fixed level.
Finally, we discuss the benefits of our proposed methodology, and outline some possible limitations.

\section{False Discovery Proportion control by Joint Error Rate control}
\label{sec:fdp-control-by-jer}
%
The point of this article is to build an inference method that takes into account multiplicity and circularity by achieving post hoc FDP control, while maintaining satisfactory statistical power.

\subsection{Notation}
We denote by $m$ the number of hypotheses, i.e. the number of voxels under consideration (typically spanning a given brain template).
In the context of fMRI, $m$ generally ranges from $50,000$ to $400,000$.
We denote the set of true null hypotheses (voxels with no effect) by ${H}_{0}$, and by $m_0 = |{H}_{0}|$ its cardinal.
Given a set of $m$ $p$-values associated to each  hypothesis, we denote by $p_{(k: m)}$ the $k^{th}$ one in ascending order.
For a set $S$ of hypotheses of interest (i.e. the set of voxels in a region of interest), the aim is to control the number of false positives in $S$, that is $\left|S \cap {H}_{0}\right|$, or equivalently, the corresponding proportion of false positives: $\mathrm{FDP}(S) = \left|S \cap {H}_{0}\right|/\left|S\right|$.

\subsection{Post hoc FDP control}
The most common approach to address large-scale multiplicity problems is to control the False Discovery Rate (FDR) \cite{benjamini1995controlling}. 
%
This is generally done by the Benjamini-Hochberg (BH) procedure \cite{benjamini1995controlling}, which uses different significance thresholds depending on the ranks of the $p$-values: the $k^{th}$ $p$-value is compared to $t_k^{Simes} = \alpha k / m$. 
%
%
The BH procedure controls the FDR under the PRDS (Positive Regression Dependency on a Subset) assumption~\cite{benjamini2001control}.
However, since the FDR is the \textbf{expected} FDP, FDR control is a weak statistical guarantee on the \textbf{actual} FDP \cite{korn2004controlling}.
This can be problematic when the FDP distribution has heavy tails, which can happen when the tested hypotheses are dependent. 
In such cases, the FDR might be controlled while FDP quantiles diverge (see Figure 2.1 in \cite{neuvial:tel-02969229}). 
We thus choose to focus on the control of the actual number (or proportion) of false positives.\\

A post hoc upper bound $V$ on the number of false positives is an integer-valued function of subsets $S$ of hypotheses that satisfies: 
\begin{linenomath}
\begin{align}
\mathbb{P}\left(\forall S, \ \left|S \cap {H}_{0}\right| \leq V(S)\right) \geq 1-\alpha \,.
\label{eq:FDPcontrol}
\end{align}
\end{linenomath}
Since $\mathrm{FDP}(S) = \left|S \cap {H}_{0}\right|/\left|S\right|$, obtaining a bound $V$ satisfying \eqref{eq:FDPcontrol} is strictly equivalent to obtaining a post hoc upper bound on the FDP. This equivalence will be used implicitly throughout the paper.\\

As described in \cite{goeman2011multiple}, the comparison between ordered $p$-values and $\left( t_k^{Simes} \right)_{k=1..m}$ can also provide post hoc FDP control.
This can be done using closed testing~\cite{marcus1976closed} combined with the following inequality:
\begin{linenomath}
\begin{align}
\mathbb{P}\left(\exists k \in\left\{1, \ldots,m_0 \right\}: p_{\left(k: m_0\right)} < t_k^{Simes} \right) \leq \alpha \,. 
\label{eq:Simes}
\end{align}
\end{linenomath}
Equation \eqref{eq:Simes} is an immediate consequence of the Simes inequality \cite{simes1986improved}, and also holds under the PRDS assumption.
The All-resolutions inference (ARI) method~\cite{rosenblatt2018all} provides a tighter post hoc bound that uses the thresholds $\alpha k /h(\alpha)$ instead of $t_k^{Simes} = \alpha k / m$ in \eqref{eq:Simes}, where $h(\alpha) \leq m$ is the so-called Hommel value~\cite{hommel1986multiple}.
$h(\alpha)$ represents an $1-\alpha$-level upper confidence bound on the number $m_0$ of true null hypotheses.\\

\subsection{Joint Error Rate}

An alternative construction of post hoc bounds has been introduced by  \cite{blanchard2020post}. 
Letting $R_{k}^{Simes}=\left\{i: p_{i} \leq t_k^{Simes} \right\}$, Equation \eqref{eq:Simes} can be written as:
\begin{linenomath}
\begin{align}
\mathbb{P}\left(\forall k,\left|R_{k}^{Simes} \cap {H}_{0}\right| \leq k-1\right) \geq 1 - \alpha\,.
\label{eq:Rk}
\end{align}
\end{linenomath}
Equation~\eqref{eq:Rk} can be interpreted as the simultaneous control of all $k-$Family-Wise Error Rate (FWER), where the $k-$FWER is the probability of obtaining at least $k$ false positives.
Each set $R_{k}^{Simes}$ yields a valid FDP upper bound over any subset $S$:

\begin{linenomath}
\begin{align*}
\left|S \cap {H}_{0}\right| 
&=\left|S \cap \overline{R_{k}^{Simes}}\cap {H}_{0}\right|+\left|S \cap R_{k}^{Simes} \cap {H}_{0}\right| \\ 
& \leq\left|S \cap \overline{R_{k}^{Simes}}\right|+\left|R_{k}^{Simes} \cap {H}_{0}\right| \\ 
& = \sum_{i \in S} 1\left\{p_{i}(X) \geq t_k^{Simes}\right\} +\left|R_{k}^{Simes} \cap {H}_{0}\right| \\ 
& \leq \sum_{i \in S} 1\left\{p_{i}(X) \geq t_k^{Simes}\right\} + k-1 \\
& =: V_k^{Simes}(S) \, ,
\end{align*}
\end{linenomath}
where the last inequality holds with probability at least $1 - \alpha$ by \eqref{eq:Rk}.\\

\begin{figure*}[t]
\centering
\includegraphics[width=0.97\linewidth]{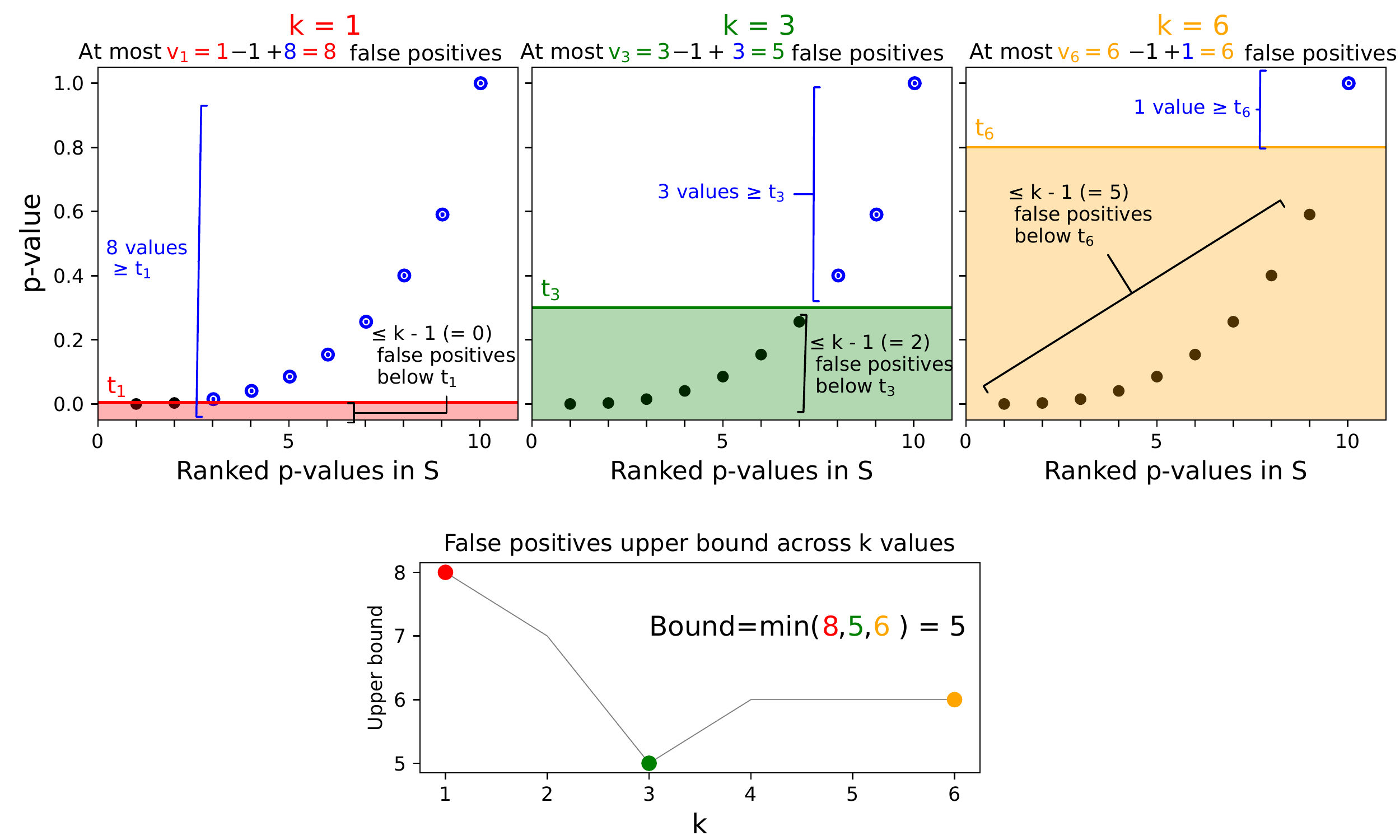}
\caption{ \textbf{Computation of the post hoc bound \eqref{eq:bound} on the number of false positives}, given a set $S$ of 10 $p$-values, using a JER controlling threshold family. Top panels: computation of $k$-th bound \add{$V_k(S) = \sum_{i \in S} 1\left\{p_{i}(X) \geq t_k\right\} + k-1 $ for 3 values of $k$, with horizontal colored lines representing the associated thresholds $t_k$.} Bottom panel: \add{The post hoc upper bound\eqref{eq:bound} corresponds to the minimum of all $V_k(S)$. In this example, the bound guarantees  that the number of false positives in the $S$ is at most 5} \del{post hoc bound computation, which corresponds to the minimum of all $k$-th bounds $V_k(S)$. In that case, we find that the number of false positives in the set of 10 $p$-values is no more than 5.} \add{with probability $> 90\%$.}}
\label{fig:bound}
\end{figure*}

The computation of $V_k^{Simes}(S)$ is illustrated in the top panels of Figure \ref{fig:bound} for $k \in \{1, 3, 6\}$.
Since \eqref{eq:Rk} holds simultaneously for all $k$, the minimum over $k$ of all $V_k^{Simes}(S)$ is a valid upper bound on the false positives in $S$ \cite{blanchard2020post}. Therefore, as illustrated in the bottom panel of Figure\ref{fig:bound}, the final post hoc FDP upper bound is $V^{Simes}(S)/|S|$, where
\begin{linenomath}
\begin{align}
V^{Simes}(S)=\min _{1 \leq k \leq|S|}\left\{\sum_{i \in S} 1\left\{p_{i}(X) \geq t_k^{Simes} \right\}+k-1\right\}\,.
\label{eq:SimesBound}
\end{align}
\end{linenomath}
As noted by \cite{blanchard2020post}, the bound \eqref{eq:SimesBound} coincides with the bound originally proposed by \cite{goeman2011multiple}. 
This can be generalized as follows by replacing $t^{Simes}:=(t_k^{Simes})_{1\leq k \leq m}$ with any threshold family $t:=(t_k)_{1 \leq k \leq k_{max}}$ corresponding to $R_{k}=\left\{i: p_{i} \leq t_k \right\}$.
\del{Here, the $k_{max}$ parameter controls the length of threshold families.}
\del{This can be exploited when the signal is a priori parsimonious, as discussed in Section~\ref{kmax}.}
%
\add{While setting $k_{\max} =m$ is natural in \eqref{eq:SimesBound}, setting  $k_{max} < m$ can be useful when $t_k$ is calibrated from the data, as discussed in the next sections.}
%
%
The Joint Error Rate (JER) of the threshold family $t$ is defined by \cite{blanchard2020post} as:
\begin{linenomath}
\begin{align}
JER(t) = \mathbb{P}\left(\exists k \in\left\{1, \ldots,k_{max} \wedge m_0 \right\}: p_{\left(k: m_0\right)} < t_k \right).
\label{eq:JER}
\end{align}
\end{linenomath}

With this notation, both Equations \ref{eq:Simes} and \ref{eq:Rk} are equivalent to JER$(t^{Simes}) \leq \alpha$. By the interpolation argument outlined above, the bound
\begin{linenomath}
\begin{align}
V^t(S)=\min _{1 \leq k \leq|S| \wedge k_{max}}\left\{\sum_{i \in S} 1\left\{p_{i}(X) \geq t_{k}\right\}+k-1\right\}
\label{eq:bound}
\end{align}
\end{linenomath}
provides a valid FDP upper bound for any threshold family $t$ such that JER$(t)\leq \alpha$~\cite{blanchard2020post}. This bound can be calculated in $O(|S|)$ for a given set $S$ using Algorithm~1 in \cite{enjalbert-courrech:powerful}.

\subsection{Tighter FDP upper bounds via randomization}

The Simes inequality \eqref{eq:Simes} ensures JER control at level at most $\alpha$ for the threshold family $(\alpha k / m)_k$. While this control is sharp for independent $p$-values, it can be conservative for positively dependent $p$-values \cite{blanchard2020post}, leading to conservative FDP bounds.
The first degree of freedom that can be leveraged to obtain tighter bounds for a given $\alpha$ is to choose the least conservative threshold family among a pre-defined set of families. 
In the case of the Simes family, this is done by choosing the threshold family $(\lambda k / m)_k$ associated to the largest $\lambda$ such that the following inequality \add{(that is, JER control)} holds:
\begin{linenomath}
\begin{align}
\mathbb{P}\left(\exists k \in\left\{1, \ldots,m_0 \right\}: p_{\left(k: m_0\right)} < \frac{\lambda k}{m} \right) \leq \alpha \,. 
\label{eq:Simes-lambda}
\end{align}
\end{linenomath}

In order to reach this goal more generally, we consider collections of threshold families called \textbf{templates} since their introduction in \cite{blanchard2020post}. 
Formally, a template is set of functions $\lambda \mapsto (t_k(\lambda))_k$ such that any fixed value of $\lambda$ corresponds to a threshold family.
For example, the Simes template corresponds to the choice: $t_k(\lambda) = \lambda k / m$ for all $k= 1 \dots m$ and $\lambda>0$.\\

The \textbf{calibration} procedure introduced in \cite{blanchard2021agnostic,blanchard2020post} uses randomization (see \cite{arlot2007some}) to obtain samples from the joint distribution of $p$-values under the null hypothesis.
As the JER \eqref{eq:JER} is a function of this distribution, these so-called randomized $p$-values allow us to select the largest possible $\lambda$ such that the JER is controlled.
Algorithm \ref{alg:perm} describes how to compute such randomized $p$-values in the case of one-sample tests, using sign-flipping \cite{roche2007,arlot2007some}. Randomized $p$-values can be obtained similarly for two-sample tests, using class label permutations instead of sign-flipping.

\begin{figure}
    \centering
    \includegraphics[scale=0.6]{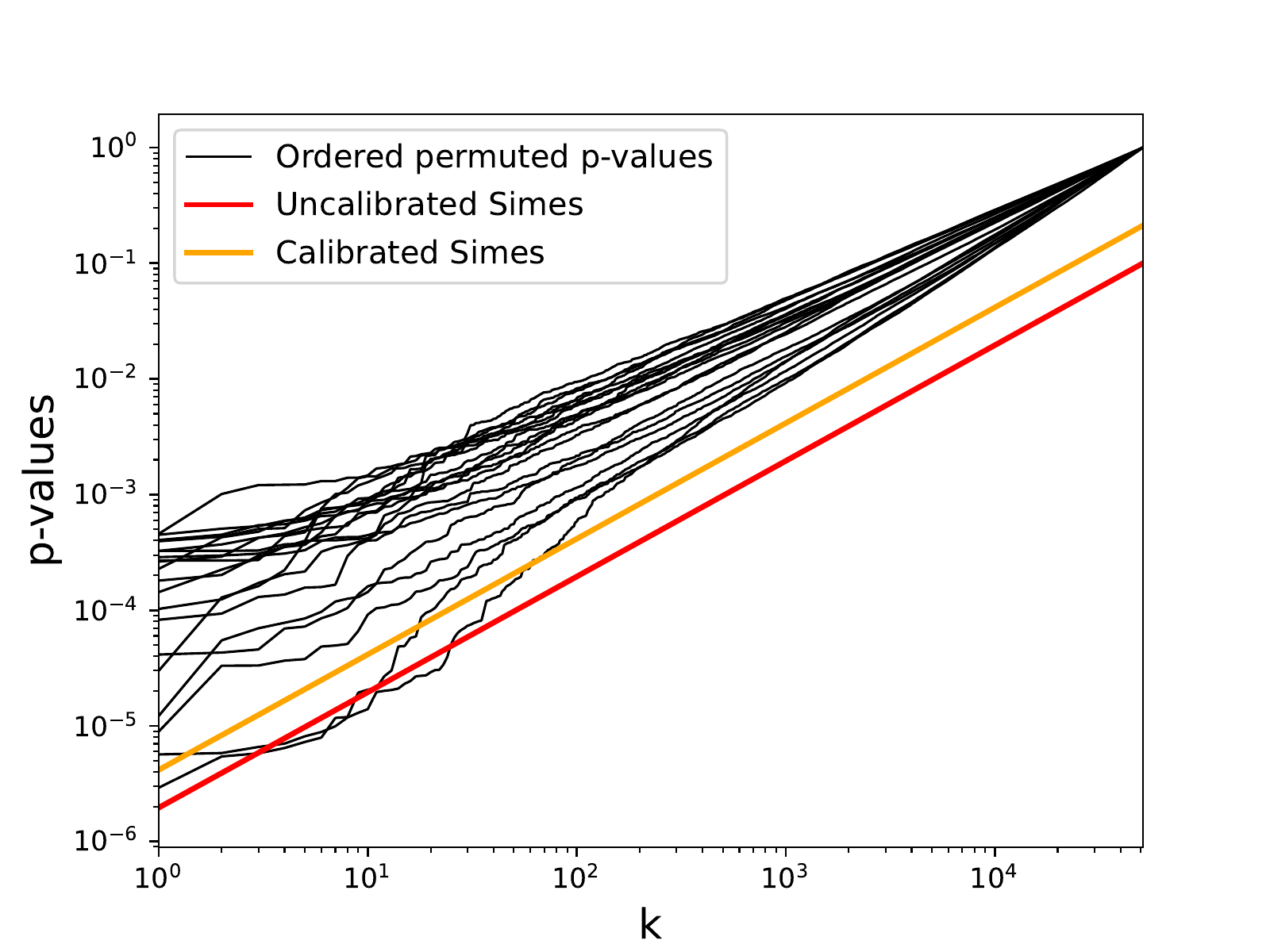}
\caption{\textbf{\del{Conservativeness of the Simes inequality and calibration} \add{Addressing the conservativeness of the Simes inequality by calibration}}. A set of 20 randomized $p$-value curves are computed on real data (black curves). Two JER controlling families at level $10\%$ are shown as colored lines. Both of them cross 2 curves (= $10\%$ of all curves) which indeed corresponds to controlling the JER at level $10\%$. The uncalibrated Simes family (in red) is conservative since it is possible to choose higher threshold families that cross the same number of black curves. The calibrated Simes family (in orange) is the least possible conservative threshold family that crosses at most 2 curves.}
    \label{fig:jer}
\end{figure}

\begin{algorithm}[H]
\begin{algorithmic}[1]
\Function{get\_randomized\_p\_values}{$X, B$}
  \State $n, p \gets$ shape(X) 
  \\ \Comment{n subjects, p voxels}
  \State pval0 $\gets$ zeros(B, p) 
  \For{$b \in [1, B]$}  
  \State \hskip 0.7cm flip $\gets$ $\operatorname{diag}$(draw\_random\_vector($\{-1, 1\}^n$)) \\ \Comment{matrix of shape (n, n)}
  \State \hskip 0.7cm $X_{flipped} = \text{flip}.X$
  \State \hskip 0.7cm pval0[b] $\gets$ one\_sample\_t\_test($X_{flipped}$, 0)
  \\ \Comment{0 = null hypothesis}
  \EndFor\\
  \State pval0 $\gets$ sort\_lines(pval0)
  \\ \Comment{Sort each vector of randomized p-values}\\
\Return pval0
\EndFunction{}
\end{algorithmic}
\caption{\textbf{Computing randomized $p$-values using sign-flipping}. For a number $B$ of sign-flips, compute $p$-values using a one-sample t-test on the flipped data $X_{flipped}$.}\label{alg:perm}
\end{algorithm}
Figure \ref{fig:jer} illustrates the conservativeness of the parametric Simes template on real data and the benefit yielded by calibration using randomized $p$-values curves. 
Choosing $\lambda>\alpha$ in \eqref{eq:SimesBound} leads to a less conservative bound.
Note that, the more dependent the data, the more the parametric Simes bound is expected to be conservative, see e.g.~\cite{blanchard2020post}.
Thus, calibration should be particularly useful for smooth data.
While the ARI procedure corresponds to using Simes inequality without calibration\footnote{Rigorously, the ARI bound corresponds to using Simes inequality with the Hommel value $h$ instead of $m$.} for JER control, calibration using the Simes template can be considered the state-of-the-art method for this problem \cite{blanchard2021agnostic,blanchard2020post}.
The bound obtained from this calibration procedure is equivalent to the bound considered in \cite{andreella2020permutation}.\\

\del{In the next section, we introduce a data-driven approach to define a candidate template.}

\section{Main contribution: data-driven templates \add{and Notip procedure}}
\label{sec:data-driven-templates}

\del{Using the above-described calibration procedure to select a threshold family based on the inference data typically yields a substantial power gain}\del{Note that the template shape is still linear as in the parametric ARI method.}

\add{The calibrated Simes family can lead to tighter post hoc bounds, yet it still relies on the Simes template, which is linear in $k$, as illustrated in Figure~\ref{fig:jer}.
Instead of only optimising $\lambda$ for a given template shape (e.g. a linear shape for the Simes template), the}
\del{The} second degree of freedom that can be exploited to achieve better statistical power while still controlling the JER is \textbf{to learn the template function, or, equivalently, its shape when displayed as a graph. }\del{, instead of only optimising $\lambda$ for a given template shape.}
Figure \ref{fig:jer} illustrates that for small $k$, permuted $p$-value curves are not exactly linear. This suggests that using a non-linear template shape could be relevant for fMRI data.
%
%
%
Several other parametric templates are considered in \cite{andreella2020permutation}, but the authors report that none of these attempts outperformed the Simes template.
An ideal template should approximately reproduce the shape of randomized $p$-values curves computed from real data.
Therefore, we propose to \textbf{learn} a template directly from the data.\\

A related idea has been explored in \cite{meinshausen2006false}.
\del{However, since the same data set was used for both the learning step and the calibration step, the method proposed in that paper}\add{However, since the method proposed in that paper does not distinguish between the learning and calibration steps}, it suffers from circularity biases, as noted by \cite[Remark 5.3]{blanchard2020post}.
Indeed, in the JER framework, the template has to be fixed a priori.\\
In order to address this issue, we propose to \emph{learn} a template from an fMRI \del{contrast} \add{dataset} that is independent from the \del{contrasts}\add{datasets} on which inference is performed.
\del{We thus assume that a training dataset is available to learn the template.
Such data can easily be obtained from public data repositories.}
\begin{figure*}
    \centering
\includegraphics[width=\linewidth]{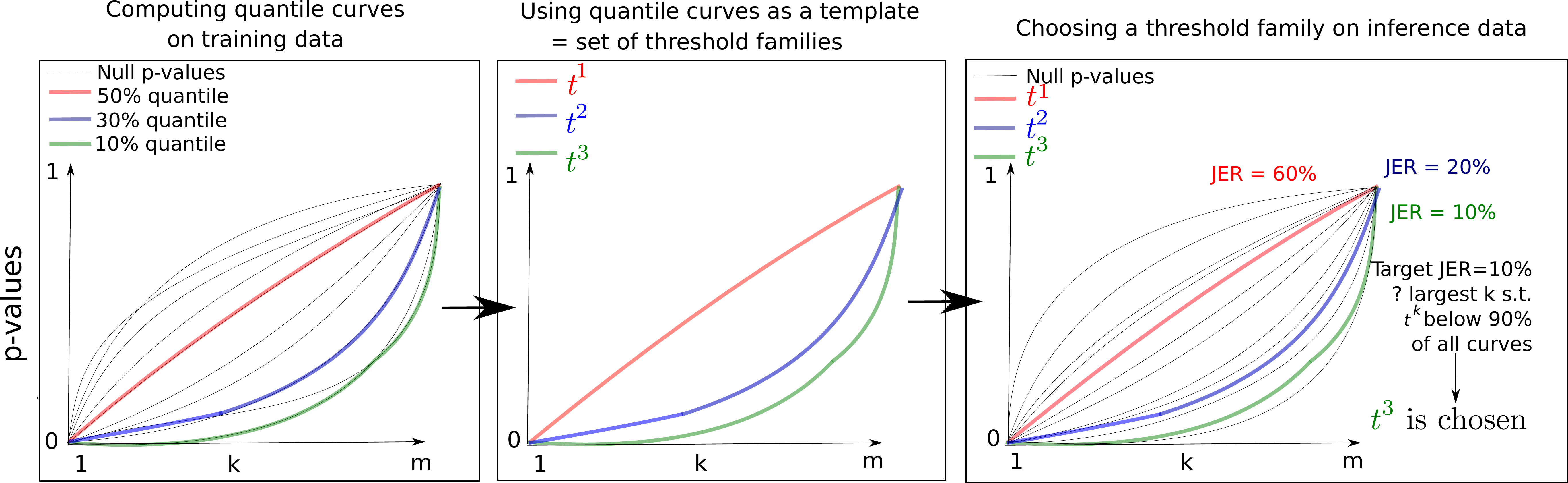}
\caption{\textbf{Learning a template from training data and using this template for calibration on inference data.} Left panel: quantiles of randomized $p$-value curves are computed on training data. Middle panel: the resulting quantile curves are used as a template (the so-called learned template). Right panel: calibration is performed on inference data using the learned template. Notice that learned templates do not have a parametric shape (contrary e.g. to the Simes template), but follow the shape of sorted null $p$-values.}
\label{fig:learned}
\end{figure*}
First, we compute $B$ randomized $p$-value curves \add{on training data} using Algorithm~\ref{alg:perm} and extract quantile curves $t^b = (t_k^b)_k$ for $b = 1 \dots B$, as shown in the left panel of Fig. \ref{fig:learned}.
These quantile curves are then viewed as a set of $B$ sorted threshold families (middle panel), which is called a \textbf{learned template}.
%
Note that it is indeed a template in the sense of \cite{blanchard2020post}, that has been discretized over a set of $B$ values.\\

After obtaining a learned template, calibration is performed on the inference data (i.e. any inference contrast) as would be done with a parametric template.
This is shown in the right panel of Figure \ref{fig:learned} and in Section~ \ref{sec:fdp-control-by-jer}. 
\add{To perform calibration, we evaluate the empirical JER of all threshold families of the learned template.} Then, we select the largest $b \in \{1, \dots B\}$ such that JER control holds on inference data for the threshold family $t^b$. \add{To avoid evaluating the JER of all threshold families, this search is done by dichotomy in practice.}\del{In practice, this is done by dichotomy.}
\add{The resulting method is called ``Notip'' for Non-parametric true discovery proportion.}
\add{
As described above, Notip requires a training dataset in order to learn the template.
An example of a template learned from a training data set is displayed in Figure \ref{fig:learnedviz}. 
Note that learned templates do not have a parametric shape, but follow the shape of randomized $p$-value curves.}\\

\add{The calibration process depends on the parameter $k_{max}$, whose choice induces the following trade-off.
On the one hand, since $JER((t^b_k)_{1 \leq k \leq K}) \leq $ \hskip 0.2cm $JER((t^b_k)_{1 \leq k \leq K'})$ for all $K \leq K'$, choosing a smaller $k_{max}$ allows calibration to choose a largest value of $b$ in the dichotomy, leading to a less conservative family.
On the other hand, a larger value for $k_{max}$ leads to more thresholds considered in the $\min$ in the bound written in Equation \ref{eq:bound}, and hence to a possibly tighter bound.
Guidelines to choose $k_{max}$ as well as an informed default choice for fMRI data are given in Section \ref{kmax}.}\\

The complete procedure is summarized in Algorithm \ref{alg:learned}, with lines 1-7 corresponding to the training step and lines 8-20 corresponding to the inference step. The latter step requires the computation of the empirical JER for a given family, which is described in Algorithm~ \ref{alg:estimateJER}.
\begin{algorithm}[H]
\caption{\textbf{Learning template on training data and calibrating on inference data.} A template is learnt by computing permuted $p$-values and extracting quantile curves. Then, this template is used to perform calibration on inference data by choosing the least conservative family of the learned template that empirically controls the JER.}\label{alg:learned}
\begin{algorithmic}[1]
\Require $X_{train}, X_{infer}, B_{train}, B_{infer}, \alpha, k_{max}$
\State $\text{pvals}_{train}$ $\gets$ get\_randomized\_p\_values($X_{train}$, $B_{train}$) \\\Comment{array of shape $(B_{train}, n_{voxels})$}\\
\Comment{lines of $\text{pvals}_{train}$ are sorted}
\For{$b \in [1, B_{train}]$}:
\State learned\_templates[$b$] $\gets $ quantiles($\text{pvals}_{train}$, $b/B_{train}$)
\EndFor
\State $\text{pvals}_{infer} \gets$ get\_randomized\_p\_values($X_{infer}$, $B_{infer}$) 
\\\Comment{vector of shape $(B_{infer}, n_{voxels})$}
\For{$b \in [1, B_{train}]$} :
\State $\widehat{JER}_b \gets \text{estimate\_jer}(\text{pvals}_{infer},
\newline
\hspace*{10.5em}   \text{learned\_templates}[b], k_{max}) $
\EndFor
\State $b_{calibrated} \gets$ $\operatorname{max}\{b \in [1, B_{train}]$ s.t. $\widehat{JER}_b \leq \alpha\}$
\\\Comment{Choose largest $b$ such that JER control holds}
\If{$b_{calibrated} = 0$} \State \Return Calibrated\_Simes
\\\Comment{No suitable learned template found}
\EndIf
\State chosen\_template $\gets$ learned\_templates[$b_{calibrated}$]\\
\Return chosen\_template
\end{algorithmic}
\end{algorithm}

\begin{algorithm}[H]
\begin{algorithmic}[1]
\caption{\textbf{JER estimation on randomized $p$-values.} The empirical JER is computed for a given template and a matrix of permuted $p$-values. This computation is directly based on Equation \ref{eq:JER}.}\label{alg:estimateJER}
\Function{estimate\_jer}{pvals, thr, $k_{max}$}
  \State ($B_{infer}$, p) $\gets$ shape(pvals)
  \State $\widehat{JER} \gets 0$
  \For{$b' \in [1, B_{infer}]$}:
    \For{$i \in [1,...,k_{max}]$}:
    \State diff[i] $\gets$ pvals$[b'][i]$ - thr$[i]$
    \\ \Comment{Check JER control at rank $i$}
    \EndFor
  \If{min(diff) $< 0$}:
  \State $\widehat{JER} \gets \widehat{JER} + 1/B_{infer}$
  \\ \Comment{Increment risk if JER control event is violated}
  \EndIf
  \EndFor\\
\Return $\widehat{JER}$
\EndFunction{}
\end{algorithmic}
\end{algorithm}

Once Algorithm \ref{alg:learned} has been run, according to \cite{blanchard2020post,blanchard2021agnostic}, the bound defined in Equation \ref{eq:bound} \del{yields} is a valid FDP upper bound.
This bound can be computed on any subset of interest $S$ in linear time in $|S|$ using Algorithm~1 in \cite{enjalbert-courrech:powerful}. 
%

\section{Experiments}
\label{sec:experiments}
\subsection{Data}

\subsubsection{FMRI data}
\label{sec:fmri-data}
To investigate the potential \del{power} gain \add{in number of detections} yielded by using data-driven templates, we performed experiments on an fMRI dataset, collection 1952 \cite{varoquaux2018atlases} of the Neurovault database (\url{http://neurovault.org/collections/1952}).
This dataset is an aggregation of 20 different fMRI studies, consisting of statistical maps obtained at the individual level for a large set of contrasts.
These images have been preprocessed using the procedure described in \cite{varoquaux2018atlases}. 
In particular, they have been spatially normalized to MNI space using SPM12 software, and resampled to 3mm isotropic resolution.
In the present case, the inference question concerns one-sample tests in group analyses, i.e. identifying what brain regions show a significant increase of activity for the contrast of interest, as opposed to the baseline,
across participants.
The group-level statistic and associated $p$-value are obtained through a one-sample t-test on the individual z-maps.\\

\del{Since collection 1952 only contains elementary \emph{'versus baseline'} contrasts, we had to find relevant pairs of contrasts to obtain meaningful inference examples.
Such 'versus baseline' contrasts contain a massive amount of non-specific signal, hence we pair them with control contrasts.}
\add{
Collection 1952 only contains elementary \emph{'versus baseline'} contrasts, with a massive amount of non-specific signal.
In order to obtain meaningful inference examples, we paired them with control contrasts.}
A typical interesting contrast pair is "words vs baseline" vs "face vs baseline"; by subtracting these two contrasts, we obtain the more relevant "words vs face" contrast, which aims at uncovering brain regions with significantly higher or lower signal for word images than for face images stimuli.\\

To obtain consistent results, we excluded contrasts with too few subjects and/or trivial signal. The resulting list of 36 contrast pairs is given in Table \ref{tab:contrasts}.\\ 
 
In order to use data-driven templates on fMRI data, we have to choose a training set beforehand, on which we learn a template once and for all. \add{The variability of the Notip method with regards to the choice of the training set is studied in Section 4.3. For the rest of the experiments, we use a single training set.}
Although \del{choosing} \add{learning} a different template for each contrast pair would produce statistically valid inference, the computational cost would be high and this would lead to a loss in generality (i.e. the user would have to learn a template per inference contrast pair, instead of doing it once). 
For these experiments, we choose \del{a training pair of contrasts to learn the data-driven template} \add{for training data a pair of contrasts} with 113 subjects and 51199 voxels smoothed using $FWHM$ (full width at half-maximum) $= 4mm$ and \add{at least} $2\%$ of active voxels (\del{as estimated using }\add{with probability $\geq 95\%$ according to} ARI). 
This is the pair of contrasts with the lowest proportion of active voxels that we could find among contrast pairs with at least 100 subjects. 
This choice \add{is referred to as the optimal template in the rest of the paper. It} is explained in the Discussion.
This template is learnt using $B_{train} = 10,000$ permutations and \add{we choose} $k_{max} = 1,000 \simeq \lfloor m/50 \rfloor$ for reasons detailed in Section~\ref{kmax}. Note that we also apply the same choice of $k_{max}$ when using the Simes template, so that both templates are compared on a fair basis.\\

\subsubsection{Synthetic data}
\label{sec:synthetic-data}
\add{For some of the experiments described below, we have generated simulated data using the pyrft package:  \url{https://github.com/sjdavenport/pyrft}. This package allows generates smooth noisy random fields that resemble fMRI data. In this controlled setup, the ground truth is known. An example of such simulated data can be found in Section \ref{simexample}. The simulation setting is the following, with $\pi_0$ the proportion of null voxels: $\alpha = 0.05, \text{ } \pi_0 = 0.9,\text{ } FWHM = 4mm, \text{ } n_{train} = 100, \text{ } n_{infer} = 50, \text{ } q = 0.1, \text{ } B_{train} = B_{infer} = 1000$. }
\subsubsection{Code to reproduce the experiments}
Data manipulation is mostly performed through Nilearn v0.9.0, nibabel v.3.1.1. The proposed statistical methods are implemented in the sanssouci package: \url{https://github.com/pneuvial/sanssouci.python}.
The experiments presented in this section can be reproduced using the code at\del{ the following address}: \url{https://github.com/alexblnn/Notip}. This repository contains a script per experiment.\\

The analysis \del{work} we performed on this data can be divided into \del{4} \add{6} main experiments that are detailed in the rest of this section.\\

\subsection{\texorpdfstring{\del{Detection rate} Variation \add{of the number of detections} for all three methods}{Variation of the number of detections for different template types}}
\label{experiment1}

To compare different choices of templates and investigate whether data-driven templates yield a \del{detection rate} gain \add{in number of detections} over existing methods, we compute the size of the largest possible region that satisfies a target error control for each choice of template on the 36 chosen contrast pairs.
This is typically the type of inference that users \del{perform with} \add{aim for when applying} FDR controlling procedures such as the Benjamini-Hochberg procedure. \add{We denote by $S_t$ the largest region (i.e. subset of voxels) such that its FDP upper bound is smaller than some user-defined value $q \in [0,1]$, called the FDP budget. It corresponds to the maximum FDP that one is willing to tolerate in a given region.} Formally, we solve the following optimisation problem for any template $t$:
\begin{linenomath}
\begin{equation}\label{eq:St}
|S_t| = \max_{S} \left|S\right| \quad
\textrm{   s.t.} \quad \frac{V^{t}_{\alpha}(S)}{\left|S\right|} \leq q\,, \end{equation}\end{linenomath}
where $V^{t}_{\alpha}(S)/|S|$ is the upper bound on the FDP at risk level $\alpha$ computed on $S$ using the template $t$.
\add{By construction of the bound~\eqref{eq:bound}, the solution of \eqref{eq:St} is a $p$-value level set, of the form $\{i/p_i \leq \tau\}$ for some $\tau$ \cite[Section 7.4]{blanchard2020post}. As such,} \del{Note that} $|S_t|$ can be obtained in linear time in $m$ using Algorithm 1 in \cite{enjalbert-courrech:powerful}.\\

Then, we compute the relative size difference of $S_t$ for all possible pairs of \del{templates} \add{methods}. Formally, \textbf{the variation of the number of detections} \del{\textbf{detection rate variation}} between the learned template (i.e., the Notip procedure) and the calibrated Simes template is defined as:\begin{linenomath}$$\frac{|S_{Learned}|- |S_{Simes}|}{|S_{Simes}|}$$\end{linenomath}

The calibration procedure on any a priori fixed template controls the JER \cite{blanchard2020post,blanchard2021agnostic}.
Therefore, it makes sense to compare the \add{number of detections} \del{detection rate} obtained by different template choices (i.e. ARI, calibrated Simes and learned template) \del{by comparing the number of detections} for a given error control \add{$1-\alpha$}.
We compare the number of detections for several values of $q$, the FDP budget, for a given risk $\alpha = 5\%$.\\

\add{We also perform the same experiment on the simulated data described in Section~\ref{sec:synthetic-data}. In this case, since the ground truth is known, we can compare the empirical True Positive Rate (TPR) of all three methods. 
This quantity represents the proportion of true signal recovered by the template $t$ for the region $S_t$ defined in \eqref{eq:St}. 
Formally, we defined the TPR in $S_t$ as the ratio of the lower bound on the true positives in $S_t$ to the number of truly activated voxels in $S_t$:}
  \begin{equation*}
\textrm{TPR}(S_t)= \frac{|S_t|-V^{t}_{\alpha}(S_t)}{|H_1|}.
\end{equation*}
\add{Where $|H_1|$ corresponds to the number of truly activated voxels. As such, TPR$(S_t)$ is an empirical measure of power for the template $t$.}

\subsection{Comparison with FDR control}

The above experiment on \add{the number of detections} \del{detection rate variation} leads to a natural comparison\del{with}  \add{based on the ``BH region'', that is } the region obtained using the BH procedure that controls the FDR ( = expected FDP). 
More precisely, we compare the size of the BH region to the size of FDP controlling regions.
Conversely, we also \del{estimate} \add{compute FDP upper bounds} on the BH region. 
\add{This illustrates the difference between FDR control and FDP control with a concrete example.}\del{to evaluate how accurately the FDP is controlled using BH procedure}

\subsection{\texorpdfstring{\del{Detection rate variation} \add{Variation of the number of detections} for low sample sizes}{Variation of the number of detections for low sample sizes}}

Because of the high cost of acquisition, many fMRI datasets comprise few subjects. 
This may lead to unstable behavior and limited statistical power.
To study the impact of sample size on the inference procedure both at training and inference step, we perform two dual experiments. 
First, we compute the \add{number of detections} \del{detection rate} for the three possible \add{methods} \del{templates} as in Section \ref{experiment1} \del{the first experiment}, with the difference that the template is learned using $n_{train} = 10$ subjects instead of $n_{train} = 113$.
Second, we use the standard template with 113 subjects but this time infer on 25 pairs of fMRI contrasts with any number of subjects $n_{infer}$, varying from $n_{infer} = 8$ to $n_{infer} = 200$.

\subsection{Sensitivity to the choice of training data}
\label{sensitivity1}
\add{Since Notip requires learning a template on training data before performing inference, the choice of such data and its impact on the performance of the method is an important question. 
To assess this sensitivity quantitatively, we fix an fMRI contrast pair for inference. 
Then, we compare the number of detections for each template choice -as described in Section \ref{experiment1}- using the 36 different fMRI contrast pairs as 36 different training sets for the Notip method. It should be noted that ARI and calibrated Simes do not depend on the chosen training set; their number of detections is computed once and for all.}
\add{
These 36 fMRI contrast pairs differ in several ways such as the number of subjects, the nature of the contrasts, the fMRI study or quantity of signal. 
This allows us to evaluate the robustness of Notip to poorly matched training and inference data. 
In this experiment, we also include the optimal template choice we used for all other experiments (i.e. least amount of signal and maximum number of subjects).}

\subsection{Influence of data smoothness}

\add{Another potential source of mismatch is the smoothing done in preprocessing of fMRI data.}
\del{As in numerous statistical learning problems, the statistical properties of training and testing data ought to be well matched for the method to perform as expected.}
To assess the consequences on performance of a potential \add{smoothing} mismatch between training and inference data, we consider the case where the smoothing parameter FWHM is different in the training and inference data, using FHWM = 4mm for the training data and FWHM = 8mm for the inference data. \\

\subsection{Using Notip on a single dataset}
\label{single1}

\add{When learning a template on separate data is inconvenient, or to avoid the computational cost of learning the template, a natural idea is to use Notip on a single dataset. 
In such a setting, circularity biases may appear as in \cite{meinshausen2006false}. 
The workaround that we propose to retain valid FDP control is to perform two independent rounds of randomization - one for training and one for inference.  
While this approach is formally not covered by the theoretical framework of \cite{blanchard2020post}, we have performed experiments to assess its FDP control and power on the simulated data described in Section~\ref{sec:synthetic-data}.}

\section{Results}
\label{sec:results}
\subsection{\texorpdfstring{\del{Detection rate variation} \add{Variation of the number of detections} for different template types}{Variation of the number of detections for different template types}}

\begin{figure}[H]
\centering
\includegraphics[scale=0.6]{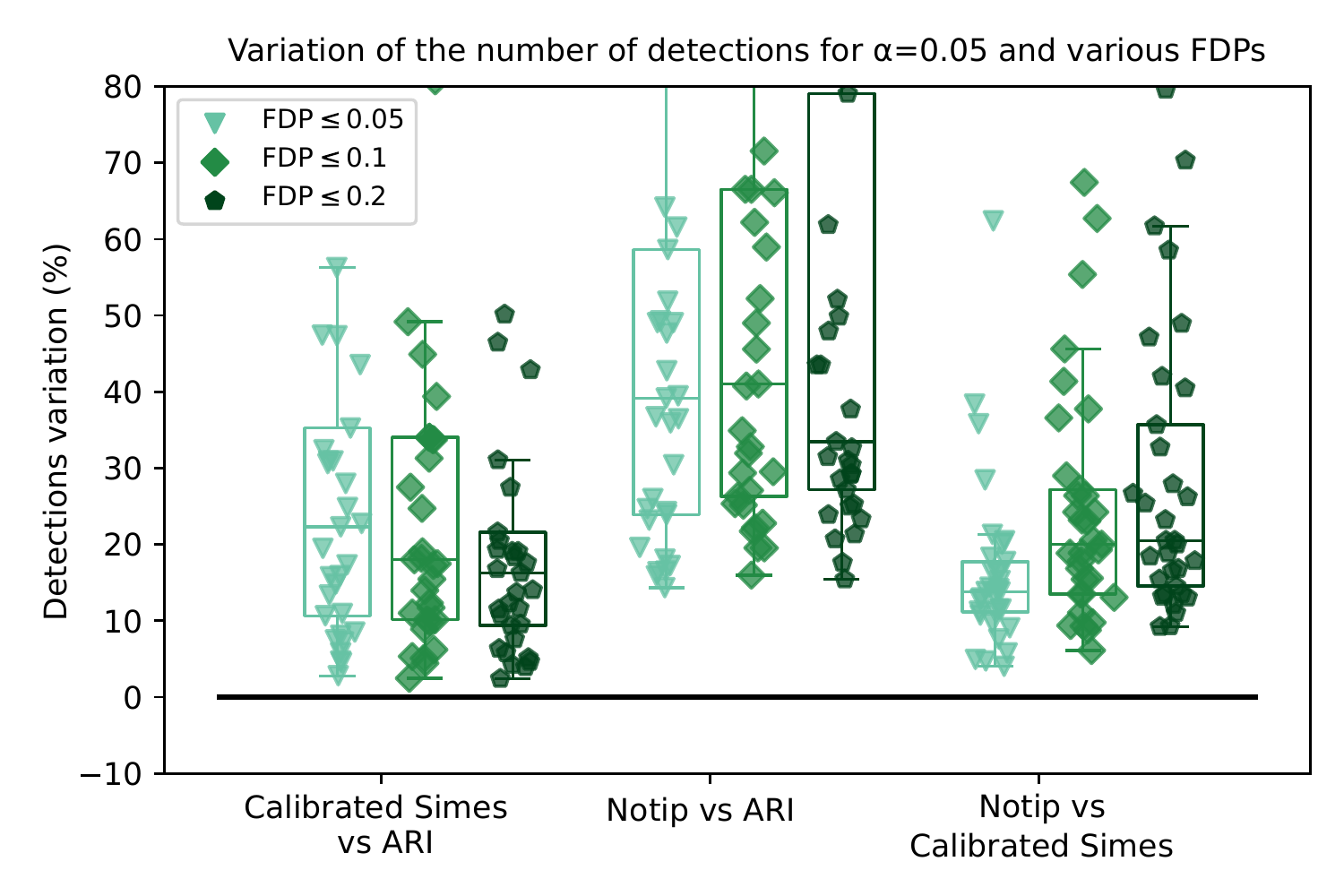}
\caption{\textbf{\del{detection rate}Comparison \add{of the number of detections} between ARI, calibrated Simes and learned templates across 36 pairs of fMRI contrasts from Neurovault collection 1952.} After learning the template on a single contrast pair (see section \ref{sec:experiments}), we perform inference on all 36 pairs. For each contrast pair, we compute the largest possible region that satisfies FDP$\leq q$ for $q \in \{0.05, 0.1, 0.2\}$ with risk level $\alpha = 0.05$.}
\label{fig:power}

\end{figure}

A comparison \add{of} the \add{number of detections} \del{detection rate} obtained for the three possible methods at hand, i.e. ARI, calibrated Simes and \add{Notip} \del{the learned template} is displayed in Figure \ref{fig:power}.
To obtain this figure, we used 36 pairs of fMRI contrasts.
The number $n_{\rm infer}$ of subjects ranged from 25 to 120 across inference contrast pairs.

\begin{figure}[H]
\centering
\includegraphics[scale=0.41]{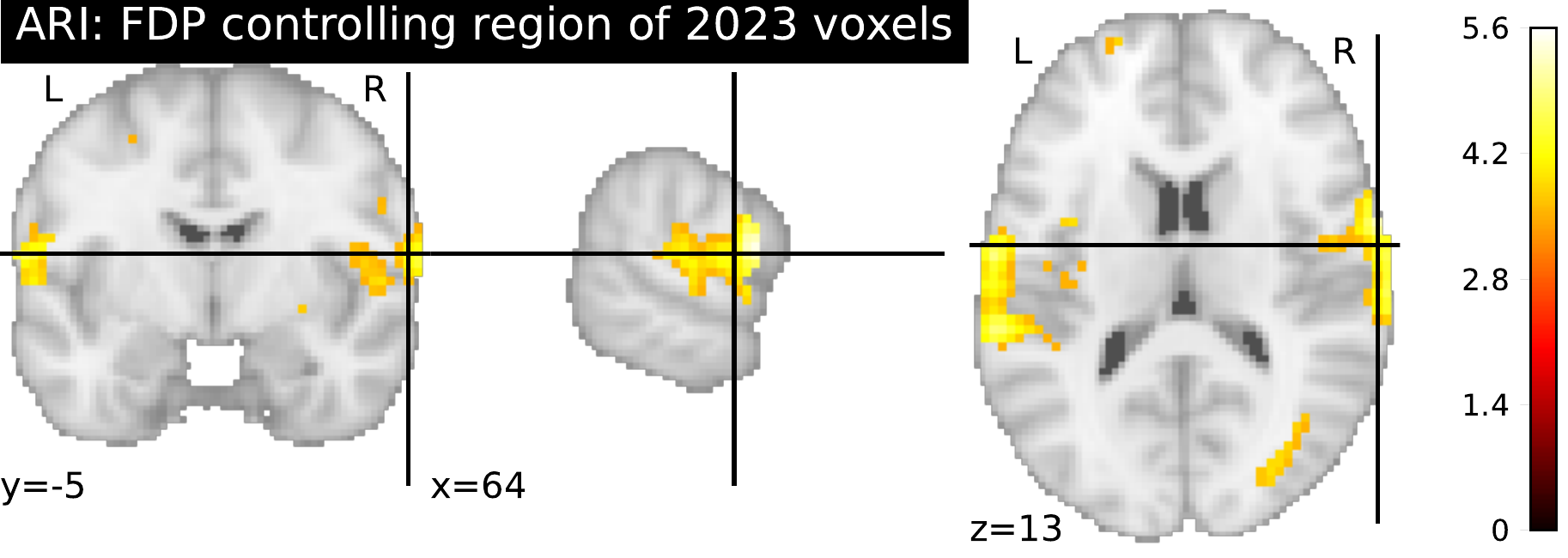}
\includegraphics[scale=0.41]{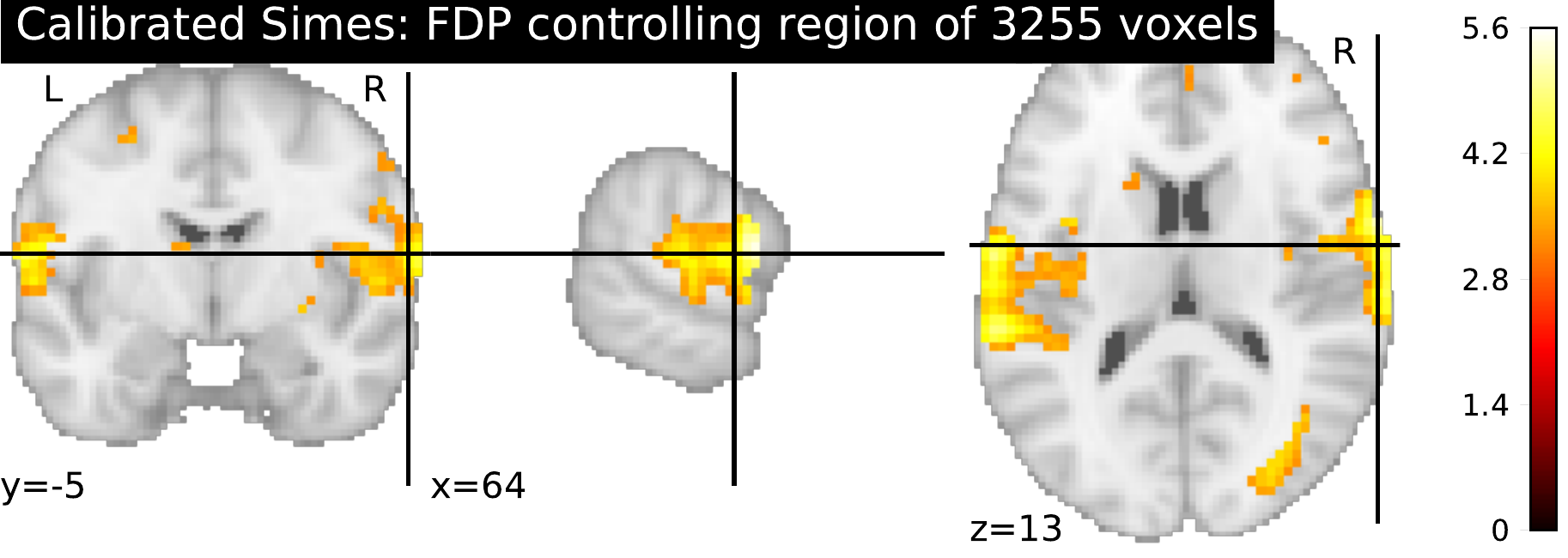}
\includegraphics[scale=0.41]{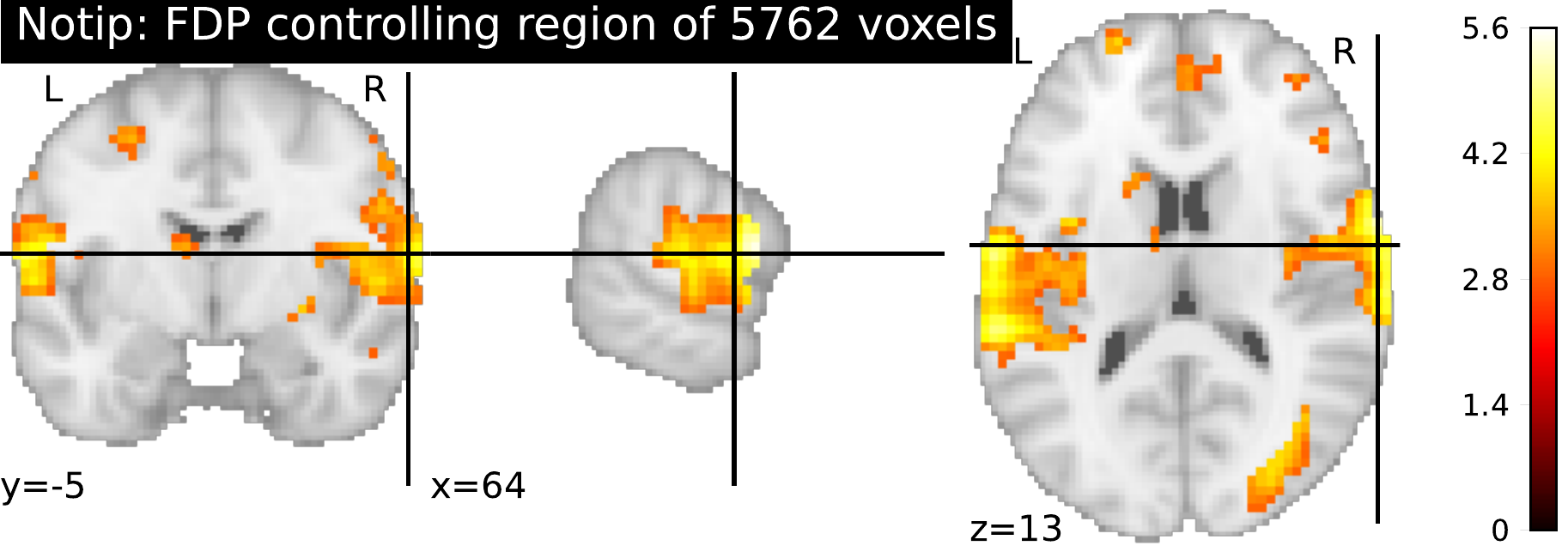}
\caption{\textbf{\del{detection rate} Comparison of the \add{number of detections} between ARI, calibrated Simes and learned template on fMRI data.} For a pair of fMRI contrasts "look negative cue" vs "look negative rating" we compute the largest possible region such that FDP $\leq 0.1$ with risk level $\alpha = 0.05$ for the three possible templates: ARI, calibrated Simes template and learned template. Notice that the \add{number of detections} \del{detection rate} is markedly higher (+ 77 \%) using the learned template compared to the calibrated Simes template.}
\label{fig:realinference}
\end{figure}

In Figure \ref{fig:power}, we notice that learned templates yield a substantial gain in \add{detections} \del{detection rate} compared to both other template choices for all \add{target FDPs}\del{requested controls}. On average, learned templates offer a $\sim \mathbf{40\%}$ \textbf{increase} in \add{detections} \del{detection rate} compared to the ARI method and a $\sim \mathbf{20\%}$ \textbf{increase} compared to calibrated Simes. \add{Gains in number of detections can vary largely across contrast pairs. This is essentially due to variance contained in the data, as all three methods exhibit similar TPR variability on simulated data (see Section \ref{powervariance})}. A concrete example of inference on fMRI data is shown in Figure \ref{fig:realinference}.

\add{We have also performed the same experiment on simulated data. In this setting, we can report the actual TPR of the methods instead of region sizes. The empirical FDP for these simulations are reported in Figure \ref{fig:FDPsim}.}

\begin{figure}[H]
\centering
\includegraphics[width=0.99\columnwidth]{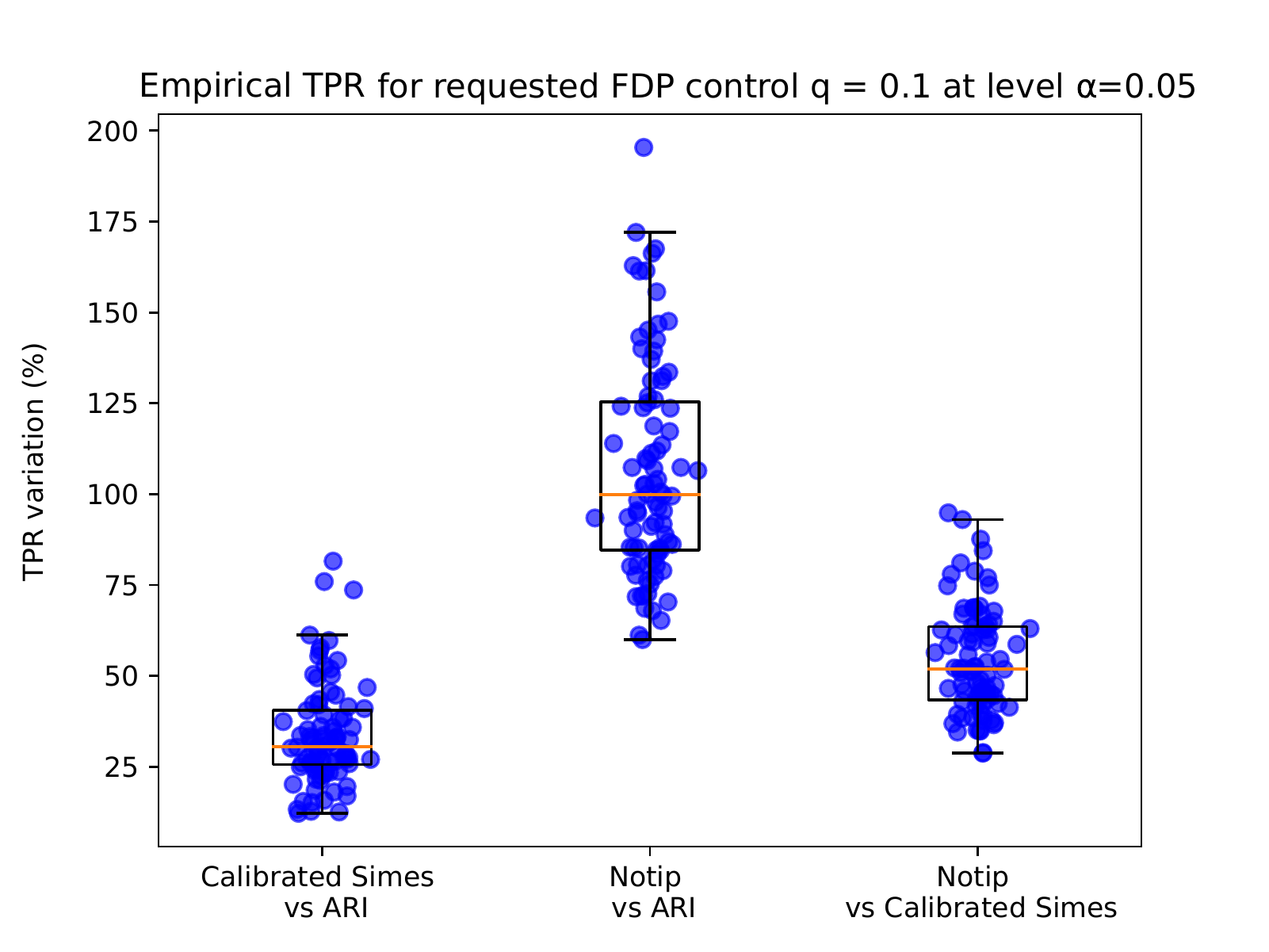}
\caption{\textbf{TPR comparison for a FDP budget $q = 0.1$ at risk level $\alpha = 0.05$}. We run 100 simulations and report the TPR. Notice that Notip  offers substantial gains in TPR compared to both ARI (100 \% on average) and calibrated Simes (50 \% on average).}
\label{fig:powersim}
\end{figure}

\add{Figure \ref{fig:powersim} illustrates the TPR gains achieved using Notip on simulated data compared to both ARI to both ARI (100 \% on average) and calibrated Simes (50 \% on average). Overall, simulations support the fact that Notip offers substantial performance gains compared to both ARI and calibrated Simes.}

\subsection{Comparison with FDR control}

\begin{figure}[H]
\centering
\includegraphics[scale=0.45]{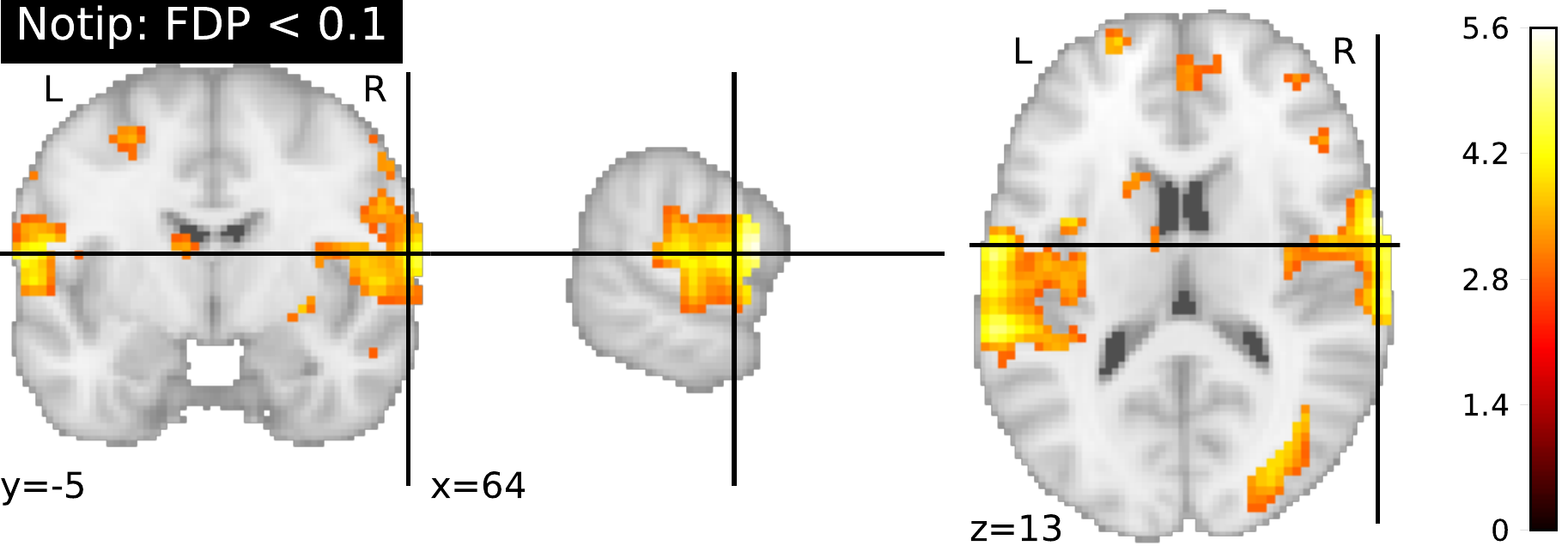}
\includegraphics[scale=0.45]{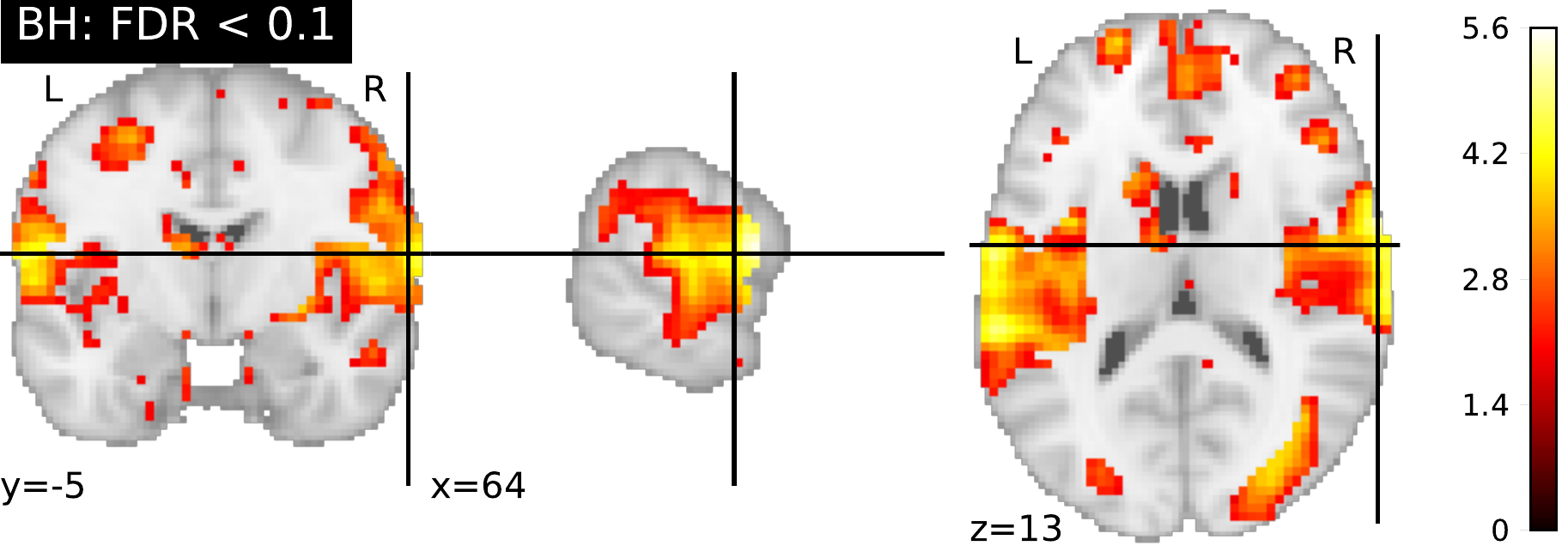}
\caption{\textbf{\del{Detection rate} Comparison \add{of the number of detections} between learned template and the BH procedure on fMRI data.} For a pair of fMRI contrasts "look negative cue" vs "look negative rating" we compute the largest possible region such that FDP $\leq 0.1$ at risk level $\alpha = 0.05 $ for the learned template and the largest possible region such that FDR $\leq 0.1$ using the BH procedure. BH region size: 13814 voxels. Learned template region size: 5762 voxels.}
\label{fig:learnedvsBH}
\end{figure}

Since FDR control is a much weaker guarantee than FDP control, it is expected that the BH procedure yields \del{a}substantially \add{more detections} \del{higher detection rate} compared to FDP controlling procedures, as seen in Figure \ref{fig:learnedvsBH}. However, FDP being the targeted guarantee, it is interesting to \del{estimate the} \add{compute} FDP \add{upper bounds} on the FDR controlling region yielded by BH. \add{Concretely, we are trying to obtain a bound on the FDP of a region that only has a guarantee on its FDR.} Table \ref{tab:EstimatedFDP} shows the \del{estimated} FDP \add{upper bounds computed} on the FDR controlling region \add{using all three possible methods.} \del{with both the calibrated Simes template and the learned template.}\\

\begin{table}[H]
    \small
    \centering
    \begin{tabular}{|c | c | c | c |}
         \hline
          & ARI & Calibrated Simes & \add{Notip} \del{Learned template} \\
         \hline
         \del{Estimated} FDP \add{Upper bound} & 61\% & 45\% & 25\%\\
         \hline
    \end{tabular}
    \caption{\textbf{\del{Estimated} FDP \add{upper bounds} on the FDR controlling region obtained using the BH procedure (at level $ q = 10 \%)$.} Notice that \add{Notip} \del{the learned template method} yields \del{more detections} \add{smaller FDP bounds} than \add {ARI} and \del{the} calibrated Simes \del{template}. \add{This upper bound remains higher than the FDR guarantee ($10 \%)$, which is more permissive by design.} \del{ but the estimated FDP remains above the FDR guarantee ($10 \%)$. In other words, in this region the FDR is controlled but likely not the FDP at level $\alpha = 0.05$ (if it were the case, we would have an estimated FDP below 10\%).}}
    \label{tab:EstimatedFDP}
\end{table}

\add{Notip leads to a less conservative FDP upper bound than ARI and calibrated Simes. However, at risk level $\alpha = 5 \%$, Notip is only able to guarantee that the FDP is less than $25\%$ while the FDR is controlled at level $10\%$. This illustrates the difference between FDR control and FDP control, the latter being less permissive by design. While the BH procedure guarantees that the \textbf{expected} FDP is below $10\%$, Notip guarantees explicitly that the \textbf{actual} FDP is below $25\%$ with high probability ($\geq 95\%$). 
It should be noted that on a single inference run, a guarantee on the \textbf{expected} FDP has no clear interpretation, whereas the guarantee on the \textbf{actual} FDP is directly interpretable.}

\subsection{\texorpdfstring{\del{Detection rate variation} \add{Variation of the number of detections} for low sample sizes}{Variation of the number of detections for low sample sizes}}

\begin{figure}[H]
\centering
\includegraphics[scale=0.6]{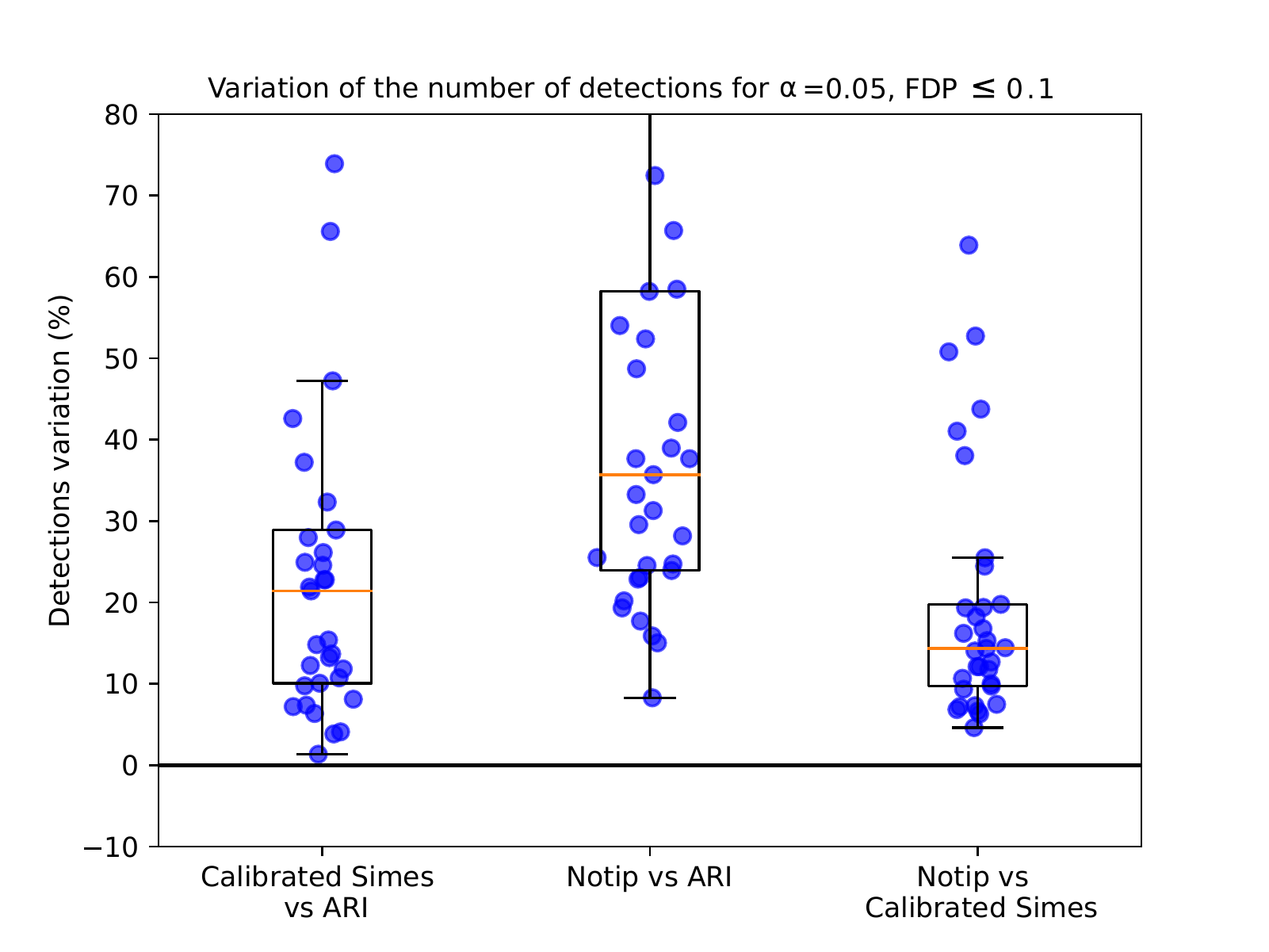}
\caption{\textbf{\del{Power} Comparison \add{of the number of detections} between ARI, calibrated Simes and a learned template using a subsampled training set.} Here, the template is learned using $n_{train} = 10$ subjects instead of $n_{train} = 113$ subjects. Learned templates still perform better than the calibrated Simes template on average, but subsampling the training set leads to a sub-optimal \add{number of detections} \del{detection rate}, compared with Figure \ref{fig:power}.}
\label{fig:capped}
\end{figure}

The above results demonstrate that data-driven templates yield consistent \del{power} gains \add{in number of detections} over existing methods that offer the same guarantees. 
In this section we investigate whether these gains subsist in sub-optimal conditions. 
Namely, when the template is learned on very few subjects or if inference is done on experiments with few subjects. 
The first point is illustrated in Figure \ref{fig:capped}.

\begin{figure}[H]
\centering
\includegraphics[scale=0.58]{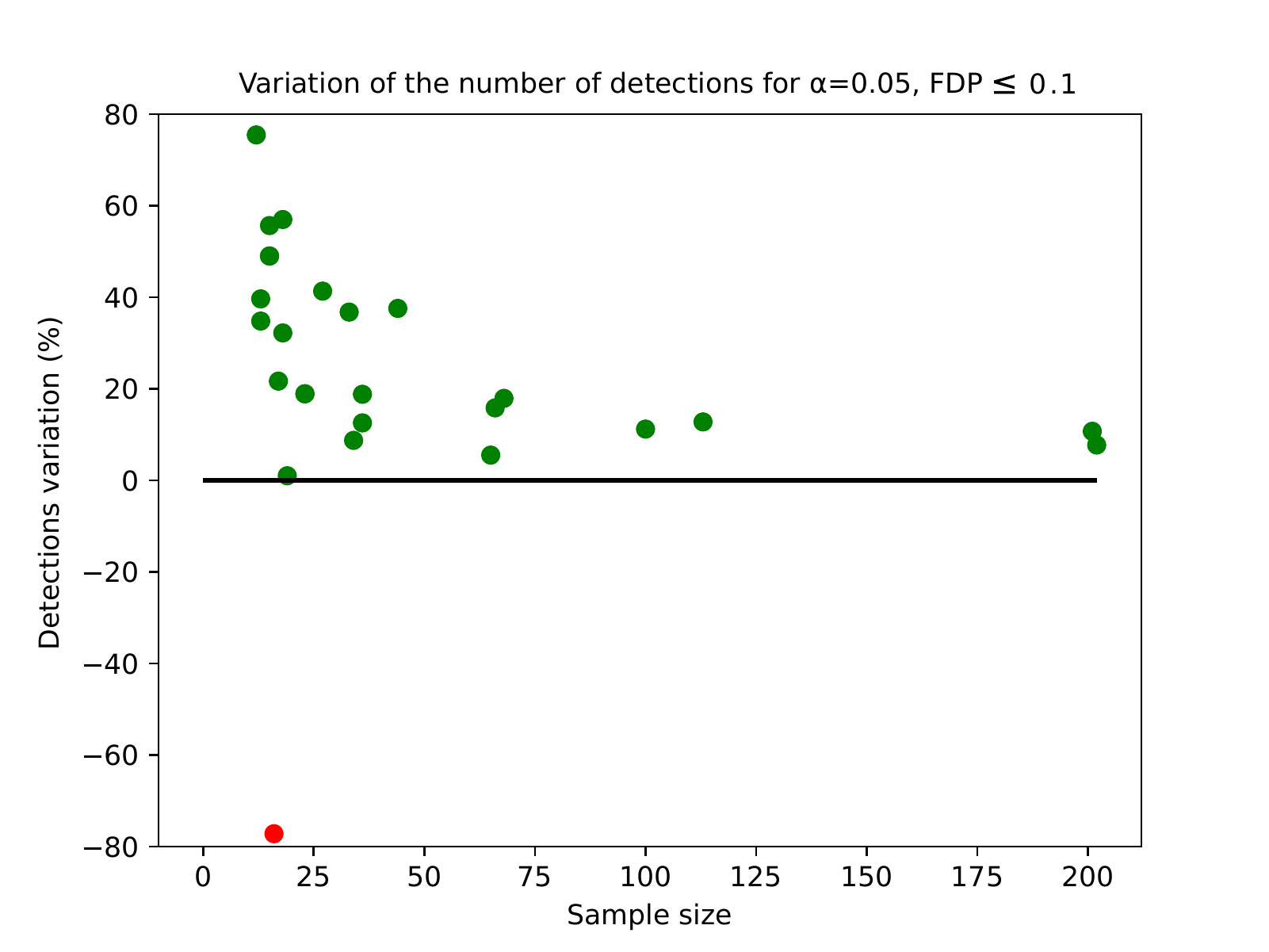}
\caption{\textbf{\del{Power} Comparison \add{of the number of detections} between learned template and calibrated Simes for many contrast pairs with a different numbers of subjects.} The \add{gains in number of detections} \del{detection rate gains}  remain consistent across datasets with different number of subjects. However, for a single dataset comprising 17 subjects, the learned template performs substantially worse than calibrated Simes.}
\label{fig:unstable}
\end{figure}

Unstable performance may occur when inferring on data with few subjects, even if the template is learned on a large number of subjects ($n_{train} = 113$ here). 
This is illustrated in Figure \ref{fig:unstable}: \del{detection rate} gains \add{in number of detections} remain consistent\add{- yet more variable for smaller sample sizes -} across datasets with different number of subjects. \add{As noted in \cite{Button2013}, high variance is unavoidable when inferring on small datasets (e.g. $n_{infer} \leq 25$).}
\del{However,}For a single dataset comprising 17 subjects, the learned template performs substantially worse than calibrated Simes.

\subsection{Sensitivity to the choice of training data}
\label{sensitivity2}
\begin{figure}[H]
\centering
\includegraphics[width=0.54\textwidth]{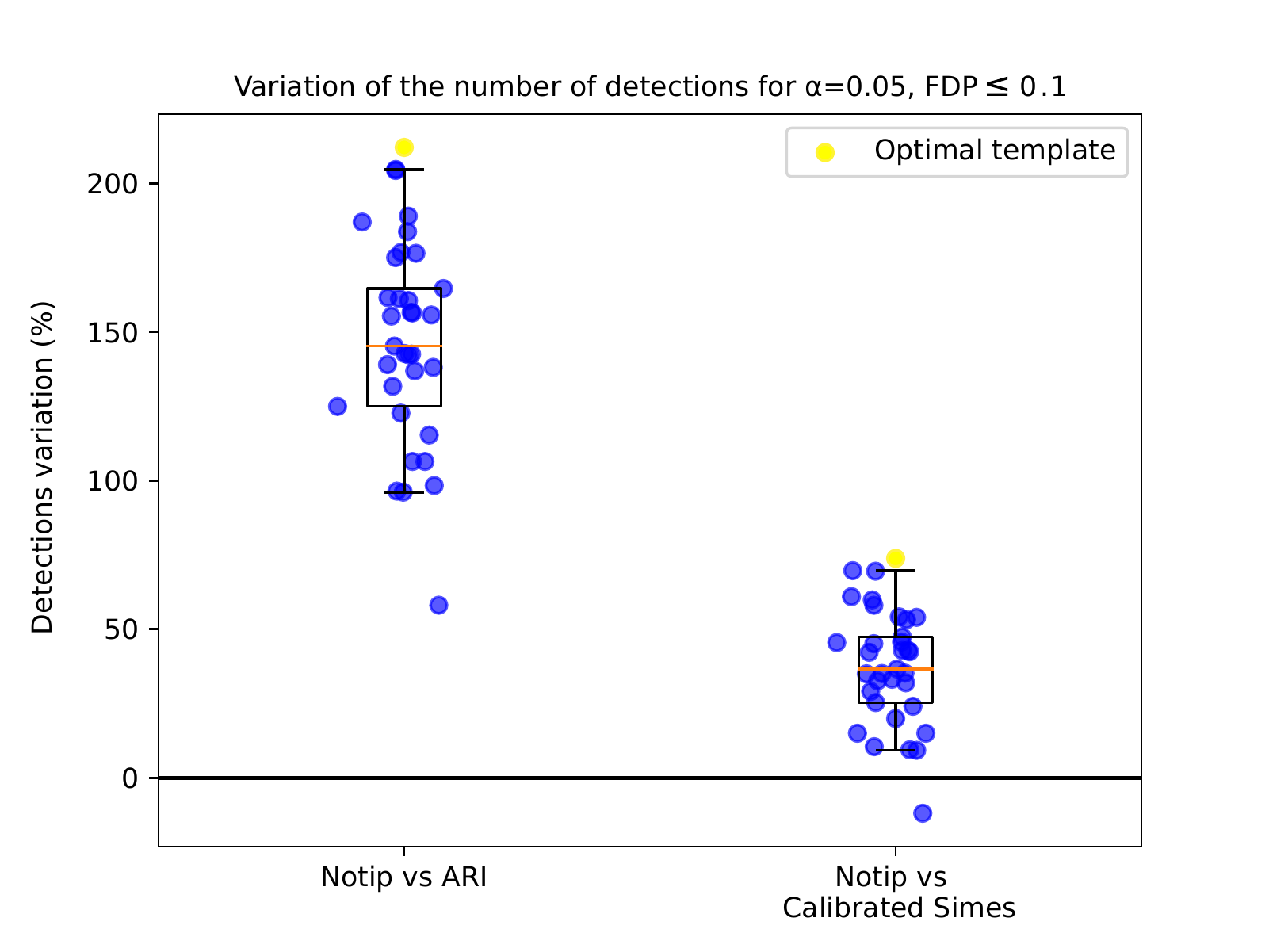}
\caption{\textbf{Variation of the number of detections using many training sets}. For a fixed contrast pair "look negative cue" vs
"look negative rating" and 36 different training contrast pairs, we compute the largest possible regions that ensure FDP $\leq 0.1$ at risk level $\alpha = 0.05$. Note that for all training contrast pairs except one, Notip performs better than calibrated Simes, with gains ranging from 10 \% to 80 \% for the optimal template choice described in Section 4.1. In the worst case, Notip performs slightly worse than calibrated Simes.}
\label{fig:notipvariance}
\end{figure}

\add{Figure \ref{fig:notipvariance} displays the variation of the number of detections made by Notip compared to ARI and calibrated Simes using 36 different training sets. All training contrast pairs except one yield more detections than calibrated Simes, with gains ranging from $10\%$ to $80\%$. This shows that the Notip procedure is robust to poorly matched training and inference data, since contrast pairs considered for training vary along many dimensions: number of subjects, nature of contrasts, fMRI study, quantity of signal... In the worst possible case, Notip performs marginally worse than calibrated Simes. Also note that the optimal template used in all other experiments (corresponding to the template learned from the training data with minimal signal and maximum number of subjects as described in Section~\ref{sec:fmri-data}) outperforms all other choices.}

\subsection{Influence of data smoothness}

We have seen in Figure \ref{fig:notipvariance} that Notip is robust to mismatches of training and inference data across different dimensions (number of subjects, quantity of signal...). We now examine the robustness of Notip with regards to a mismatch of the smoothing parameter between training and inference data.

\begin{figure}[H]
\centering
\includegraphics[scale=0.6]{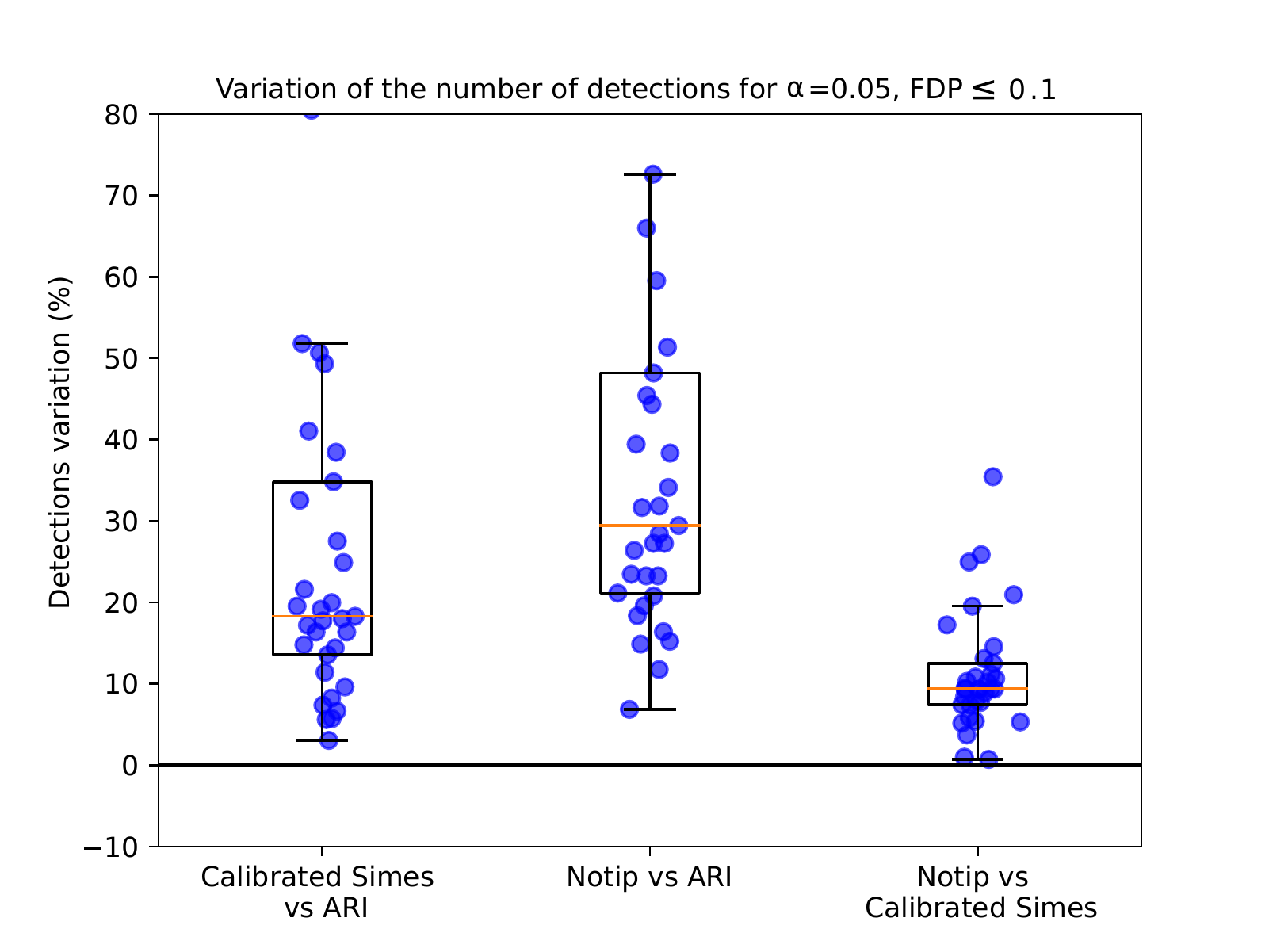}
\caption{\textbf{An example of mismatch between the smoothing factors of training and inference data.}
  After learning the template on a single contrast pair (see Section \ref{sec:experiments}) with smoothing full width at half maximum (FWHM) 4mm, we perform inference on all 36 pairs smoothed with FWHM 8mm.
  For each contrast pair, we compute the largest possible region that satisfies FDP control at level 0.1 with risk level $\alpha = 0.05$. The learned template still performs marginally better than calibrated Simes in this case, but gains are substantially lower in this regime.}
\label{fig:smoothing}

\end{figure}

Figure \ref{fig:smoothing} shows that the smoothing parameter of the training data and the inference data \del{have to} \add{should} be matched \add{for optimal performance}. 
Otherwise performance gains relative to the calibrated Simes method are reduced, albeit still positive.

\subsection{Using Notip on a single dataset}
\label{single2}
\add{To assess whether using Notip with the same dataset for training and inference controls the FDP, and whether it yields performance gains compared to ARI and calibrated Simes, we performed 1000 simulations. For each of these runs, we report the empirical FDP and TPR of all three methods.}
\begin{figure}[H]
\centering
\includegraphics[width=0.54\textwidth]{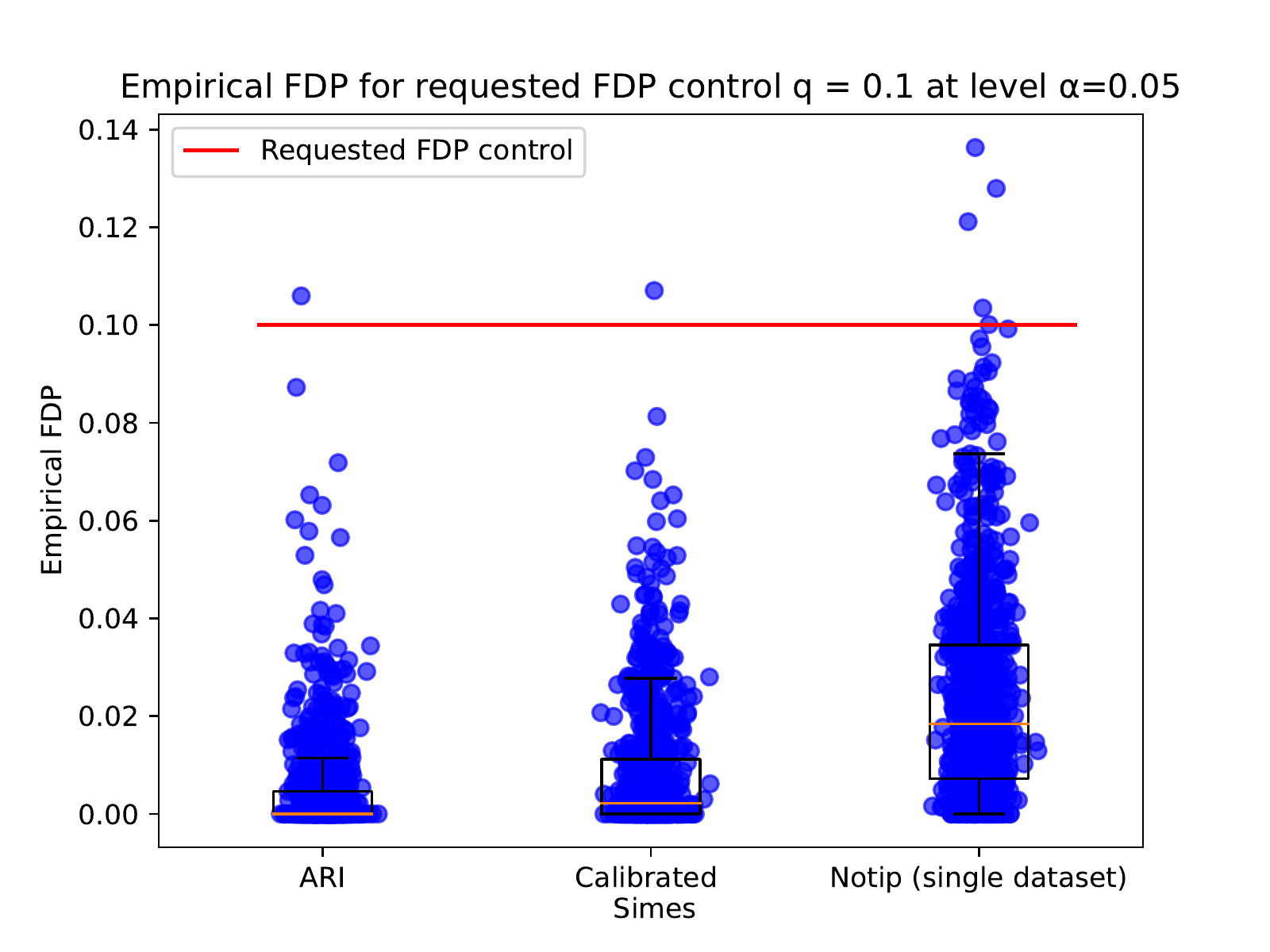}

\caption{\textbf{False Discovery Proportion achieved for a FDP budget $q = 0.1$ with risk level $\alpha = 0.05$ using Notip on a single dataset}. We run 1000 simulations and report the empirical FDP for each one. Notice that Notip (single dataset) controls the FDP at level $\alpha = 0.05$ since FDP control is violated for 5 runs, i.e. $0.5 \% < 5\%$ of all simuations. As expected, ARI and calibrated Simes also control the FDP.}
\label{fig:FDPsingledata}
\end{figure}

\add{Notice that as seen in Figure \ref{fig:FDPsingledata}, Notip (single dataset) indeed controls the FDP, as only 5 points are above the red line - i.e. the FDP was above the \add{budget} $q=0.1$ in $0.5\%$ of experiments ($< \alpha = 5 \%$). As expected, ARI and calibrated Simes control the FDP more conservatively.}

\begin{figure}[H]
\centering
\includegraphics[width=0.54\textwidth]{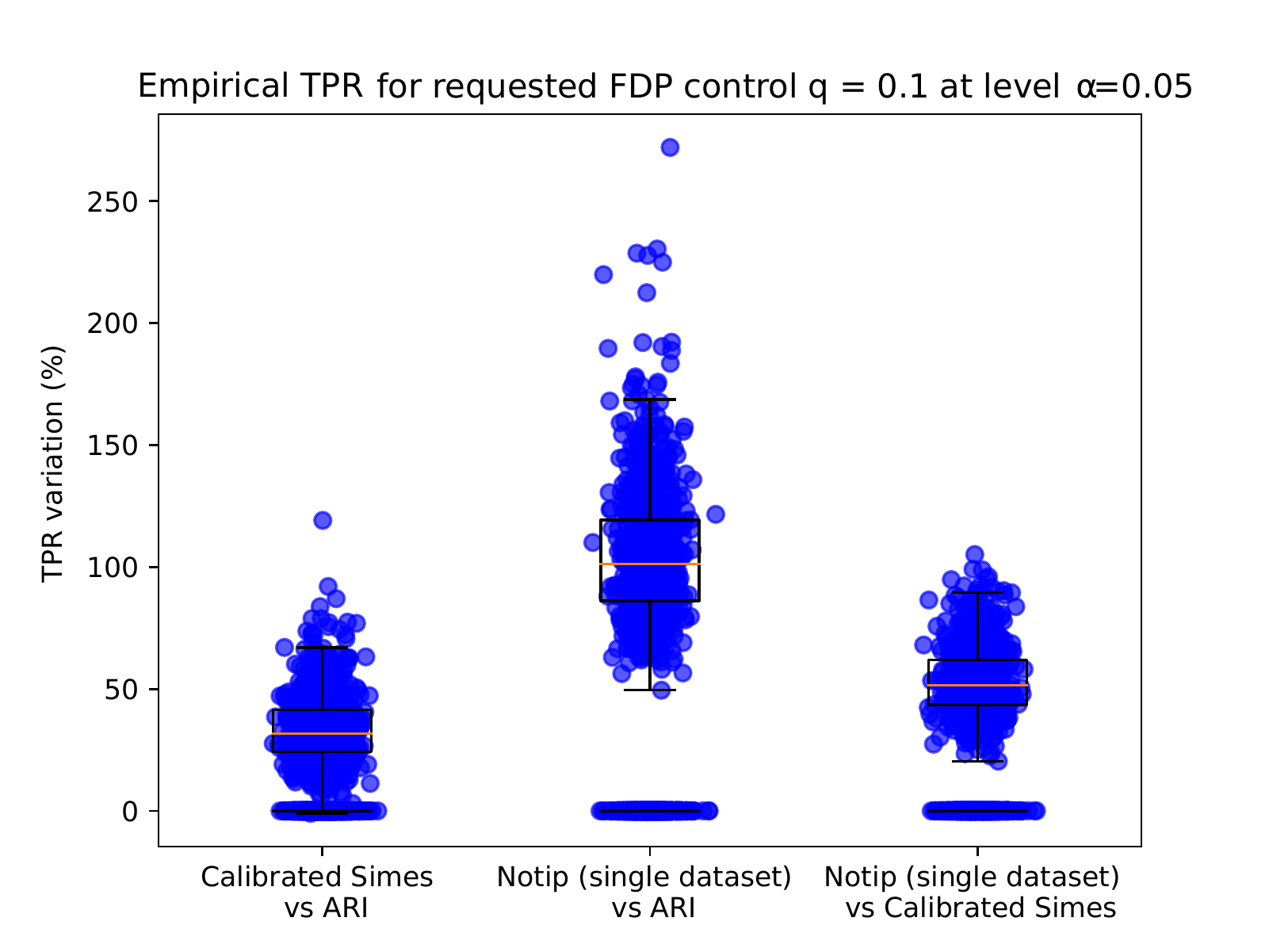}
\caption{\textbf{TPR comparison for an FDP budget $q = 0.1$ with risk level $\alpha = 0.05$ using Notip on a single dataset}. We run 1000 simulations and report the empirical TPR for each one. Notice that Notip (single dataset) offers substantial performance gains compared to both ARI (100 \% on average) and calibrated Simes (50 \% on average).}
\label{fig:POWERsingledata}
\end{figure}

\add{As seen in Figure \ref{fig:POWERsingledata}, Notip (single dataset) yields substantial performance gains compared to ARI and calibrated Simes: $50\%$ on average compared to calibrated Simes, and $100 \%$ on average compared to ARI. These gains are comparable to those obtained using the classical Notip method on simulated data (see Figure \ref{fig:powersim}).}

\section{Discussion}

In this paper, we have proposed the Notip procedure, that allows users to \del{estimate} \add{obtain statistical guarantees on} the proportion of truly activated voxels in any given cluster.
There are at least two ways to perform inference on fMRI data using this procedure. 
First, one can threshold a statistical map to obtain the largest possible region that satisfies a requested FDP control.
Second, users can also \del{estimate the TDP} \add{obtain an upper bound on the FDP, or, equivalently, a lower bound on the TDP} in any cluster of interest (see an example in Section \ref{tdpclusters}).\\

This type of analysis is meant to mitigate the arbitrariness of cluster-forming thresholds in cluster-level inference, which remains a popular framework.
The underlying observation is that estimates computed on these clusters may be plagued by circularity.\\

We have introduced a data-driven approach to obtain valid post hoc FDP control, thus achieving this goal. Moreover, controlling the FDP is a substantially more \add{informative} \del{precise} guarantee than controlling the FDR, its expected value. \del{While FDP control comes at an unavoidable power cost compared to FDR control, }We show that our procedure yields a higher \add{number of detections} \del{detection rate} than existing methods that offer the same statistical guarantees, namely ARI and calibrated Simes. We could go further by applying a step-down procedure as described in \cite{blanchard2020post}, but the gains are expected to be marginal \cite{enjalbert-courrech:powerful}.\\

\add{The gains in detections are maintained across practically all possible training sets, even in cases of poor matching between the training and inference datasets, as seen in Figure~\ref{fig:notipvariance}. Figure \ref{fig:smoothing} also illustrates the robustness of Notip, this time in the case of a poor match of smoothing parameters between the training and inference data. In this case, the gain in detections obtained by using the learned template is reduced, albeit still non-negligible (\add{30\% compared to ARI} and 9\% compared to calibrated Simes). We found that choosing training contrast pairs that contain a large number of subjects and low signal is optimal for performance.}

\del{However, this gain in detection rate is not systematic. First, it depends on the choice of the training set for learning the data-driven template. Interestingly, we found that certain learned templates outperformed the others in terms of detection rate.
These templates correspond to the training contrast pairs that contain a large number of subjects and low signal.}
This is coherent with intuition since a large number of subjects and minimal signal allow a more stable and accurate estimation of the distribution of $p$-values under the null.
Therefore, when selecting a template, it is useful to rely on a large-sample dataset with small signal magnitude.\\ 

\del{One should also be careful when using data-driven templates on small datasets, as their performance is sub-optimal in this setting.}
\del{In general, users should thus pay attention to the matching of training and testing data.}
\del{For instance, if the smoothing parameter is poorly matched between the training and testing data, the \del{detection rate} gain \add{in detections} obtained by using the learned template is reduced. It still remains non-negligible (\add{30\% compared to ARI} and 9\% compared to calibrated Simes).}

\del{Overall, even in deteriorated inference settings, the learned template offers substantial gains; this attests of the robustness of the Notip method.}

Notip comes with an additional computational cost compared to classical calibration using the Simes template, since we have to learn the template before inference. 
\del{However} \add{Generally, }this additional cost is acceptable in practice since learning a template on a contrast and inferring on a contrast have the same time complexity.
\add{If learning a template ex ante is inconvenient or simply impossible, for instance when users only have a single dataset at hand, we have shown numerically that it is possible (though not formally supported by the theory) to use Notip on a single dataset.}\\

We have used $10,000$ permutations for better resolution when learning the template instead of the typical $1,000$ permutations used at the inference step. Learning a template using $B_{train} = 10,000$ permutations with a standard laptop (on a single thread) takes around 7 minutes, while inferring on a contrast pair (using $B_{infer} = 1,000$ takes around 45 seconds). This can be trivially parallelized, as natively done in the implementation we propose. \\

A current limitation of the proposed method is that it only handles one-sample or two-sample designs. This method could be extended to multivariate linear models in future work.\\

The idea of learning templates is not specific to fMRI data and could also be used on other types of data on which the calibration procedure is useful such as genomics~\cite{enjalbert-courrech:powerful}.\\

We have achieved the goal of obtaining valid post hoc FDP control - rather than FDR control, or even weaker guarantees on clusters - while maintaining \add{a} satisfactory \del{power} \add{number of detections}. 
This allows users in the brain imaging community to use more reliable inference methods that provide robust guarantees, avoiding circularity biases. 
The efforts to build such methods appear to us as important goal for the brain imaging field. The Python code used in this paper is available at \url{https://github.com/alexblnn/Notip}. This code relies on the sanssouci package available at \url{https://github.com/pneuvial/sanssouci.python}. \\

\section{Acknowledgments}

This project was funded by a UDOPIA PhD grant from Universit\'e Paris-Saclay and also supported by the FastBig ANR project (ANR-17-CE23-0011), the KARAIB AI chair (ANR-20-CHIA-0025-01) and the SansSouci ANR project (ANR-16-CE40-0019). 
The authors thank Laurent Risser and Nicolas Enjalbert-Courrech for their precious help on writing and improving the sanssouci Python code, and Samuel Davenport for useful discussions about this work.


\bibliographystyle{plain}
\bibliography{biblio}

\section{Appendix}
\subsection{Visualising learned templates}
\label{vizlearned}

\begin{figure}[H]
\centering
\includegraphics[width=0.54\textwidth]{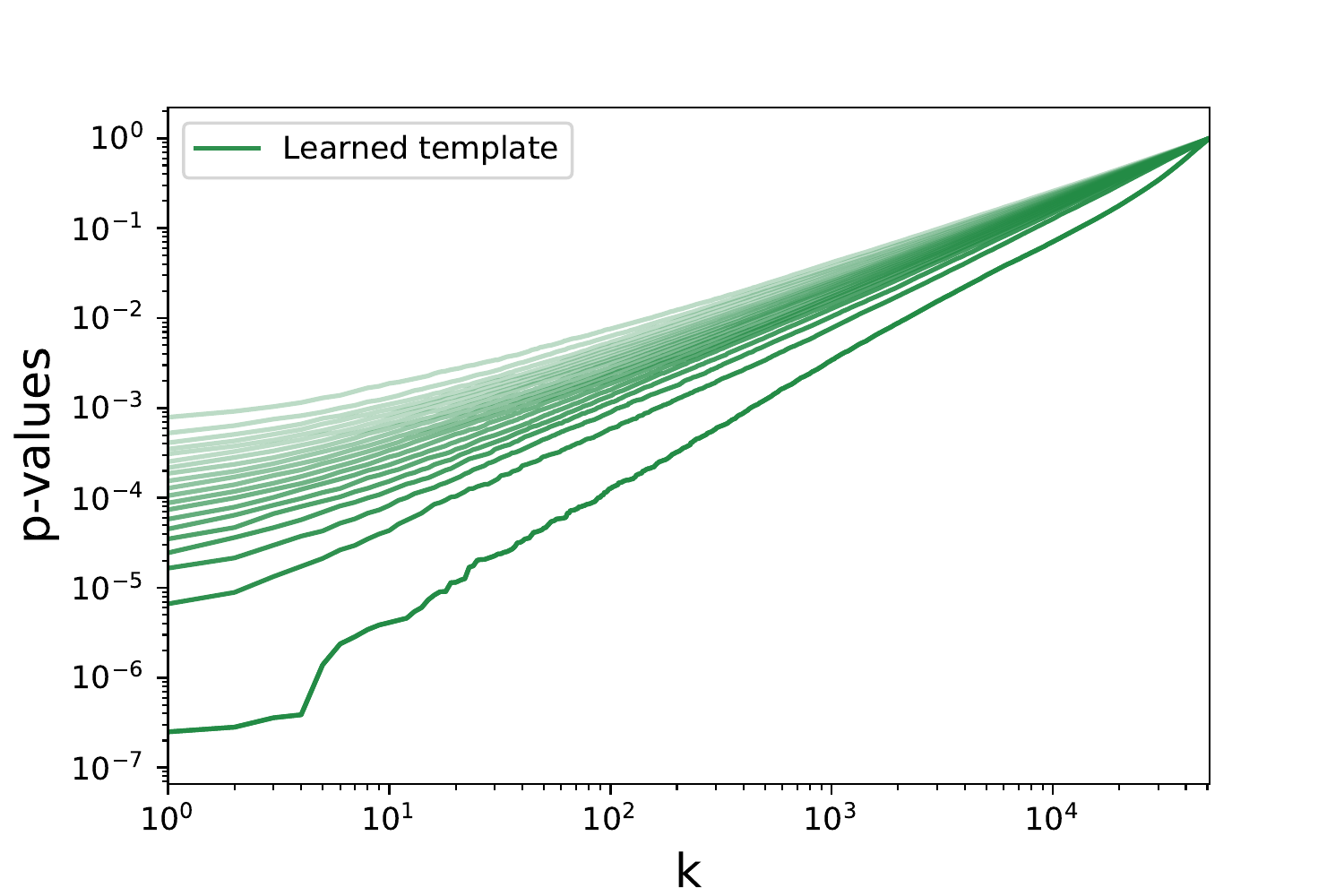}
\caption{\add{\textbf{Visualising a learned template in log-log scale}. This learned template was computed using $B = 1000$ on the same contrast pair used as training data for the experiments described in Sections \ref{sec:experiments} and \ref{sec:results}. A set of $20$ curves are retained for clarity. The curves are increasingly lighter for higher order quantiles.}}
\label{fig:learnedviz}
\end{figure}

\add{Figure~\ref{fig:learnedviz} shows an example of a learned template computed on the contrast pair used as a training set in all the experiments of the paper, using $B = 1000$. We retain a set of $20$ quantile curves for clarity - as $1, 000$ or $10, 000$ curves would not be suitable for visualization. Notice that all threshold families of the learned template are non-linear for small values of $k$, i.e. $k \leq 1000$. By definition, these curves are similar to the randomized $p$-values curves displayed in Figure \ref{fig:jer}.}

\subsection{Choice of $k_{max}$}
\label{kmax}

The post hoc bound \eqref{eq:bound} is valid for any value of the parameter $k_{max}$, provided that this parameter is chosen \textit{a priori} and not after data analysis \cite{blanchard2020post}. 
While some guidelines are given in the Discussion of \cite{blanchard2020post}, the choice of $k_{max}$ remains an open question.
Equation \ref{eq:bound} may be written as follows:
\begin{linenomath}
\begin{align}
  \label{eq:Vk}
  V(S)=\min _{1 \leq k \leq|S| \wedge k_{max}} V_k(S)\,,
\end{align}
\end{linenomath}
where $V_k(S) = \sum_{i \in S} 1\left\{p_{i}(X) \geq t_k\right\} + k-1$.
Each $V_k(S)$ is itself an upper bound on the number of false positives in $S$.
The choice of $k_{max}$ implies a tradeoff. 
On the one hand, large values of $k_{max}$ can seem advantageous because the minimum in \eqref{eq:Vk} is taken on a larger set of values of $k$. On the other hand, when the thresholds $t_k$ are obtained by calibration --- as in \cite{blanchard2020post} or in the present paper, a smaller $k_{max}$ leads to larger values of $(t_k)$ for a given $k$, and thus to a tighter  bound $V_k$.
%
%
%
Noting that $V_k(S) \geq k-1$, the values of $k$ such that $k > q |S|$ will yield $V_k(S)/|S| \geq q$ for any $S$. 
Therefore, these values of $k$ are useless for obtaining a FDP bound less than $q$.
This motivates a choice of $k_{max}$ of the form
\begin{linenomath}
\begin{align}
    \label{eq:k-max}
    k_{max} &= q_{max} |S_{max}|\,,
\end{align}
\end{linenomath}
where $q_{max}$ is the maximum proportion of false positives that can be tolerated by users and $|S_{max}|$ is the size of the largest set of voxels of interest.\\

In practice, the regions of interest are those in which a \textbf{high proportion of activated voxels} can be guaranteed. To be conservative, we set $q_{max} = 0.5$, which simply means that we are not interested in guaranteeing that the FDP is less than $q$ for $q \geq 0.5$.
In the case of fMRI, one is generally interested in sparse activation extent, as widespread effects are by definition not informative on the specific involvement of brain regions in the contrast of interest. 
As a default choice, we observe that most fMRI contrasts studied in the literature lead to less of $5\%$ of the image domain to be declared activate, which amounts to setting $|S_{max}| = 0.05 m$.

\begin{figure*}
\centering
\includegraphics[scale=0.6]{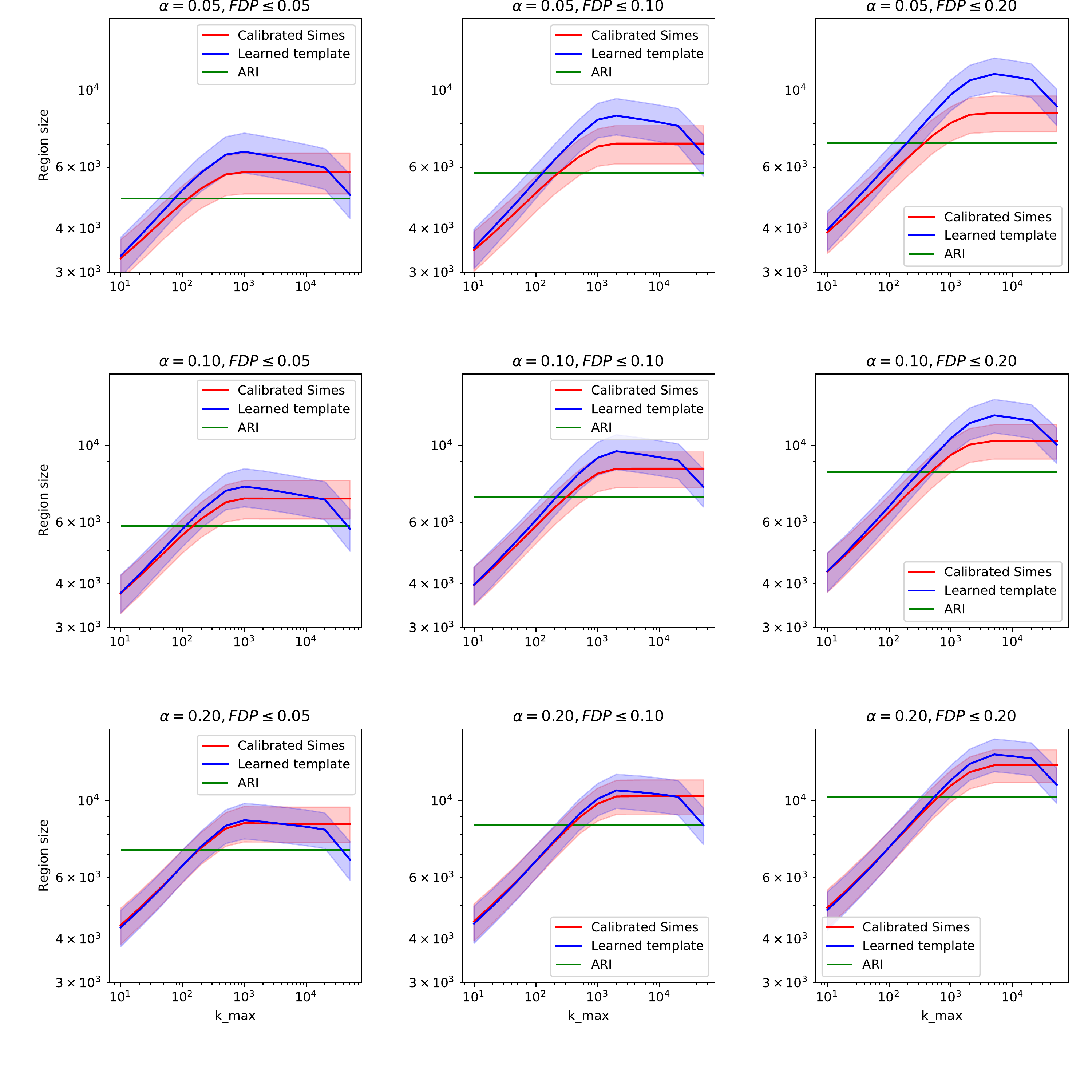}
\caption{\textbf{\del{Power} Comparison \add{of the number of detections} between learned template and calibrated Simes for various $k_{max}$ values with $5\%$ error bands in log-log scale.} Notice that the chosen $k_{max}$ largely influences the maximum size of the FDP controlling region for the learned template.}
\label{fig:kmax}
\end{figure*}

\begin{figure}[H]
\centering
\includegraphics[width=0.54\textwidth]{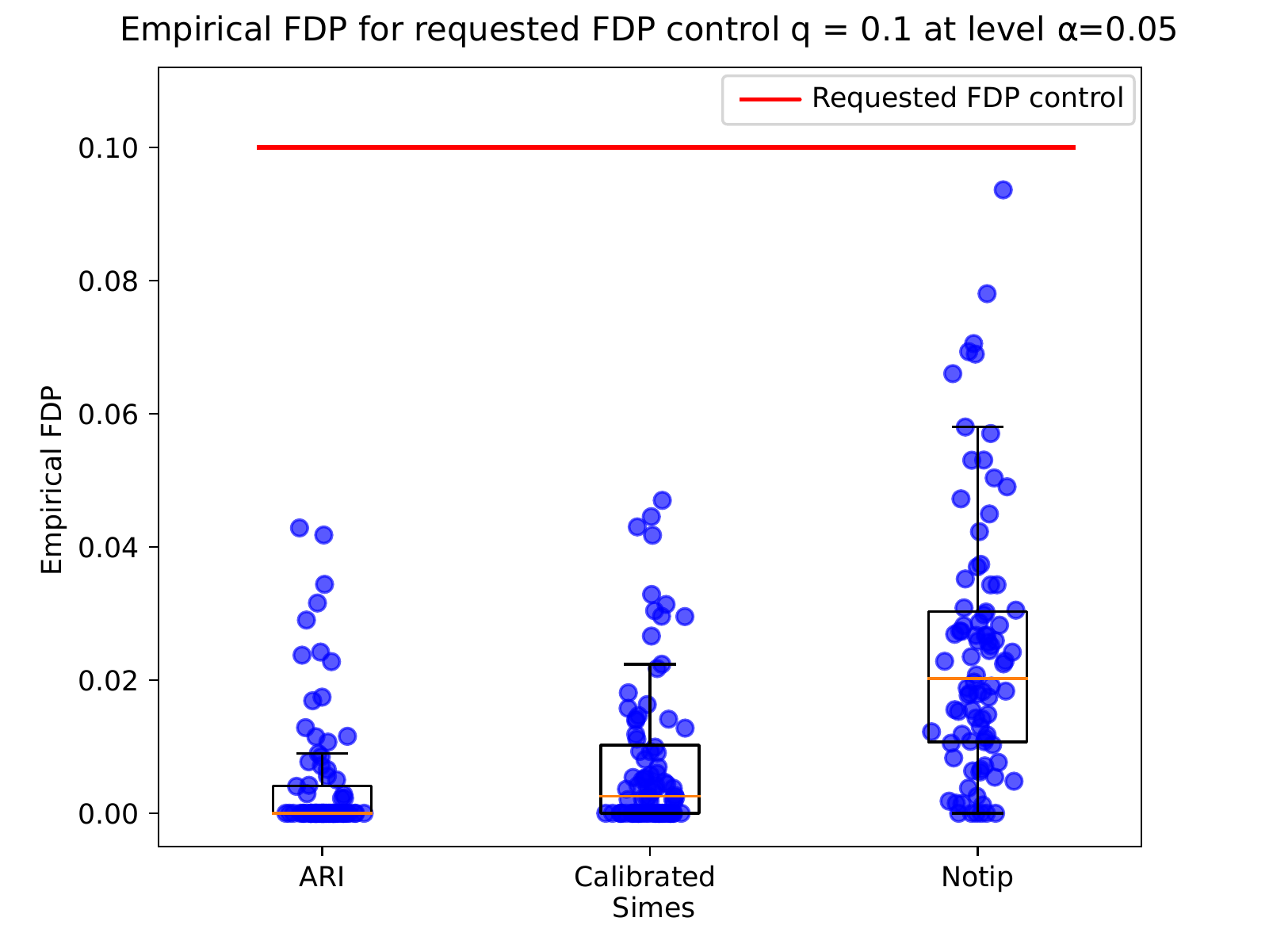}
\caption{\textbf{False Discovery Proportion achieved for a FDP budget $q = 0.1$ with risk level $\alpha = 0.05$}. We run 100 simulations and report the empirical FDP for each one. All three methods control the FDP, but Notip is less conservative than ARI and Calibrated Simes.}
\label{fig:FDPsim}
\end{figure}

Finally, a reasonable choice seems to be $k_{max} = 0.5 * 0.05 m = 0.025 m$. 
In the context of the experiments we described where $m \simeq 50,000$, we settle for simplicity on using $k_{max} = 0.02m = 1,000$. \add{This is the default value of $k_{max}$ in the implementation we propose.}
To illustrate the effect of the choice of $k_{max}$ we display the \add{variation of the number of detections} \del{detection rate variations} of all three methods on 36 fMRI datasets across 9 different inference settings for varying $k_{max}$ in Figure \ref{fig:kmax}.
Except for extremely small or large values of $k_{max}$ Notip is at worst slightly sub-optimal and $k_{max} = 1,000$ is a safe default.\\

As noted in \cite{blanchard2020post}, no choice of $k_{max}$ uniformly outperforms others. For example, the above choice, which is motivated by the \emph{prior}: "$|S_{max}| = 0.05 m$", may be poorly adapted in situations where very large regions are considered.

\subsection{FDP control on simulated data}

\add{In section 4.2 we report the empirical TPR for experiments on simulated data, for which the ground truth is known. We also compute the FDP for each simulation run to verify that, as expected, Notip indeed controls the FDP.}

\subsection{Variability of Notip}
\label{powervariance}
\add{We have observed relatively high variability in number of detections when comparing Notip to ARI and calibrated Simes in Figure \ref{fig:power}. One may wonder whether this variability is inherent to the Notip procedure or stems from the data. To assess this, we report the empirical TPR of each method (rather than the 3 pairwise comparisons) on simulated data, in the same setup as in Figure \ref{fig:powersim}.}

\begin{figure}[H]
\centering
\includegraphics[width=0.54\textwidth]{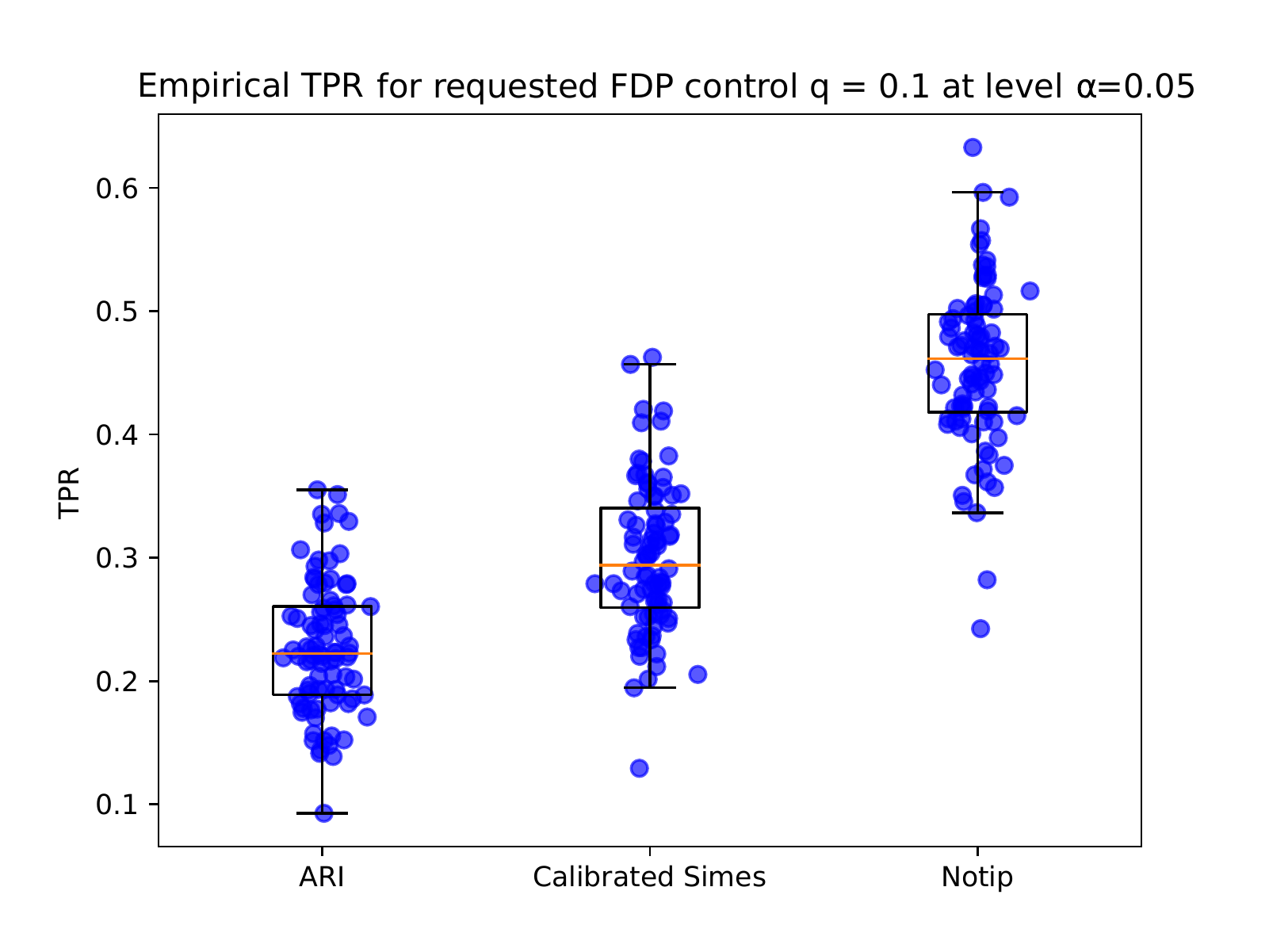}
\caption{\textbf{TPR comparison for an FDP budget $q = 0.1$ with risk level $\alpha = 0.05$}. We run 100 simulations and report the empirical TPR for each one. Notice that the variability of performance is similar for all three methods.}
\label{fig:powervariance}
\end{figure}

\add{Figure \ref{fig:powervariance} indicates that all three methods exhibit similar performance variability on simulated data. This suggests that the variability observed in Figure \ref{fig:power} is due to the data itself rather than to the Notip method.}

\subsection{\texorpdfstring{TDP \del{estimation} \add{lower bounds} on clusters}{TDP lower bounds on clusters}}
\label{tdpclusters}
\del{While we chose to compare this method's power in the standard inference setting of fMRI - i.e. find the largest possible region that satisfies a certain control - the method also yields valid inference on the TDP of data-driven clusters. This is illustrated in Table \ref{tab:clusters}.} 
\add{Throughout the paper, we chose to focus on FDP upper bounds - and thus on FDP controlling regions - to make Notip comparable with other methods that control the FDR or the FWER. Since Notip is a post hoc method, it can also be used for inference on data-driven clusters. In this setting, it is natural to formulate the results in terms of TDP lower bounds (obtained as 1 - FDP upper bounds), since users generally want a positive guarantee when inferring on clusters. This is illustrated in Table \ref{tab:clusters}.}
\add{Notice that Notip is able to offer less conservative guarantees on the TDP in all clusters than both ARI and calibrated Simes. In Table \ref{tab:clusters0} we retain 3 clusters among the 9 found in Table \ref{tab:clusters} for further study, i.e. changing the cluster-forming threshold to assess its impact on performances of all three methods. In Tables \ref{tab:clusters1} $(z > 2.5)$ and \ref{tab:clusters2} $(z > 3.5)$, notice that the same clusters are detected with varying sizes. The TDP guarantees remain less conservative using Notip than both ARI and calibrated Simes when the cluster-forming threshold is either lowered to 2.5 or upped to 3.5.}

\subsection{Additional details on \cite{blanchard2020post}}

\add{For self-containedness, this subsection contains additional details on FDP control via the control of the Joint Error Rate, as described in \cite{blanchard2020post}, as well as precise references to the paper's Theorems.}

\subsubsection*{JER control and FDP upper bound}

\add{JER control as defined in \ref{eq:JER} corresponds to Equation 2 of \cite{blanchard2020post}, and the post-hoc bound defined in \ref{eq:bound} corresponds to Equation 3 of \cite{blanchard2020post}}:
\begin{linenomath}
\begin{align}
JER(t) = \mathbb{P}\left(\exists k \in\left\{1, \ldots,k_{max} \wedge m_0 \right\}: p_{\left(k: m_0\right)} < t_k \right).
\tag{\ref{eq:JER}}
\end{align}
\end{linenomath}
\begin{linenomath}
\begin{align}
V^t(S)=\min _{1 \leq k \leq|S| \wedge k_{max}}\left\{\sum_{i \in S} 1\left\{p_{i}(X) \geq t_{k}\right\}+k-1\right\}
\tag{\ref{eq:bound}}
\end{align}
\end{linenomath}

\subsubsection*{Randomization}

\add{Since the distribution $p_{\left(k: m_0\right)}$ is unknown in practice, we use randomization to sample from it, as described in Section 2.4. This is detailed in Section 3 of \cite{blanchard2020post}. In one-sample designs, which is the setting considered throughout experiments, we perform random sign-flipping of samples. Let us denote $\mathcal{G}=\{-1,1\}^{n}$ the group of all possible sign-flippings $s$ of size $n$. This group acts on $X$ of shape $(n, m)$ in the following way:}
\begin{linenomath}$$
(s . X)_{i, j}=s_{i} X_{i, j}, i \in \mathbb{N}_{n}, j \in \mathbb{N}_{m} .
$$\end{linenomath}
\add{In words, for each sample index $i \in \llbracket 1, n \rrbracket$, we either sign-flip the sample $X_{i}$ or leave it untouched. Notice that under the null hypothesis, the distribution of $p_i(X)$ and of $p_i(s.X)$ are equal. Therefore, we are able to approximate the joint distribution of $\left(p_{i}(X)\right)_{i \in \mathcal{H}_{0}(P)}$ conditionally on $X$ by $\left(p_{i}(s . X)\right)_{i \in \mathcal{H}_{0}(P)}$ (see \cite{arlot2007some}).}

\subsubsection*{Calibration}

\add{The final step of the procedure is to perform calibration using the randomized $p$-values that we previously computed. Once calibration has been performed, a valid post-hoc FDP upper bound is obtained via \ref{eq:bound}. Calibration is described in Section 3 and written explicitly in Algorithms \ref{alg:learned} and \ref{alg:estimateJER}. In \cite{blanchard2020post}, calibration is described more formally, in order to obtain a proof that this procedure indeed yields JER control. For a given template $t_{k}(\cdot) \text{ with } 1 \leq k \leq \text{   } k_{max}$ and a risk level $\alpha$, the goal is to find the largest $\lambda$ such that $ JER(t_k(\lambda)) \leq \alpha $. Formally, we want to compute:}

\begin{linenomath}
$$
\lambda(\alpha)=\max \left\{\lambda \geq 0: \widehat{JER}(t) \leq \alpha \right\}
$$\end{linenomath}
With:
\begin{linenomath}
$$
\widehat{JER}(t)=\frac{1}{B} \sum_{b=1}^{B} 1\left\{\exists k \in\left\{1, \ldots, k_{\max }\right\}: p_{\left(k: m_{0}\right)}^{b}<t_{k}(\lambda)\right\}
$$\end{linenomath}

\add{This corresponds to Equation 20 of \cite{blanchard2020post}. Note that the empirical JER is computed using Algorithm \ref{alg:estimateJER} in practice. In the case of Notip, $\lambda$ takes discrete values; the maximum is computed using dichotomy as described in Section 2. Theorem 4.8 of \cite{blanchard2020post} shows that the calibration procedure indeed controls the JER, thus leading to a valid post-hoc FDP upper bound. The proof of Theorem 4.8 can be found in Section 7 of the supplementary material of \cite{blanchard2018supplement}.}

\subsection{An example of simulated data}
\label{simexample}

\begin{figure}[H]
\centering
\includegraphics[scale=0.6]{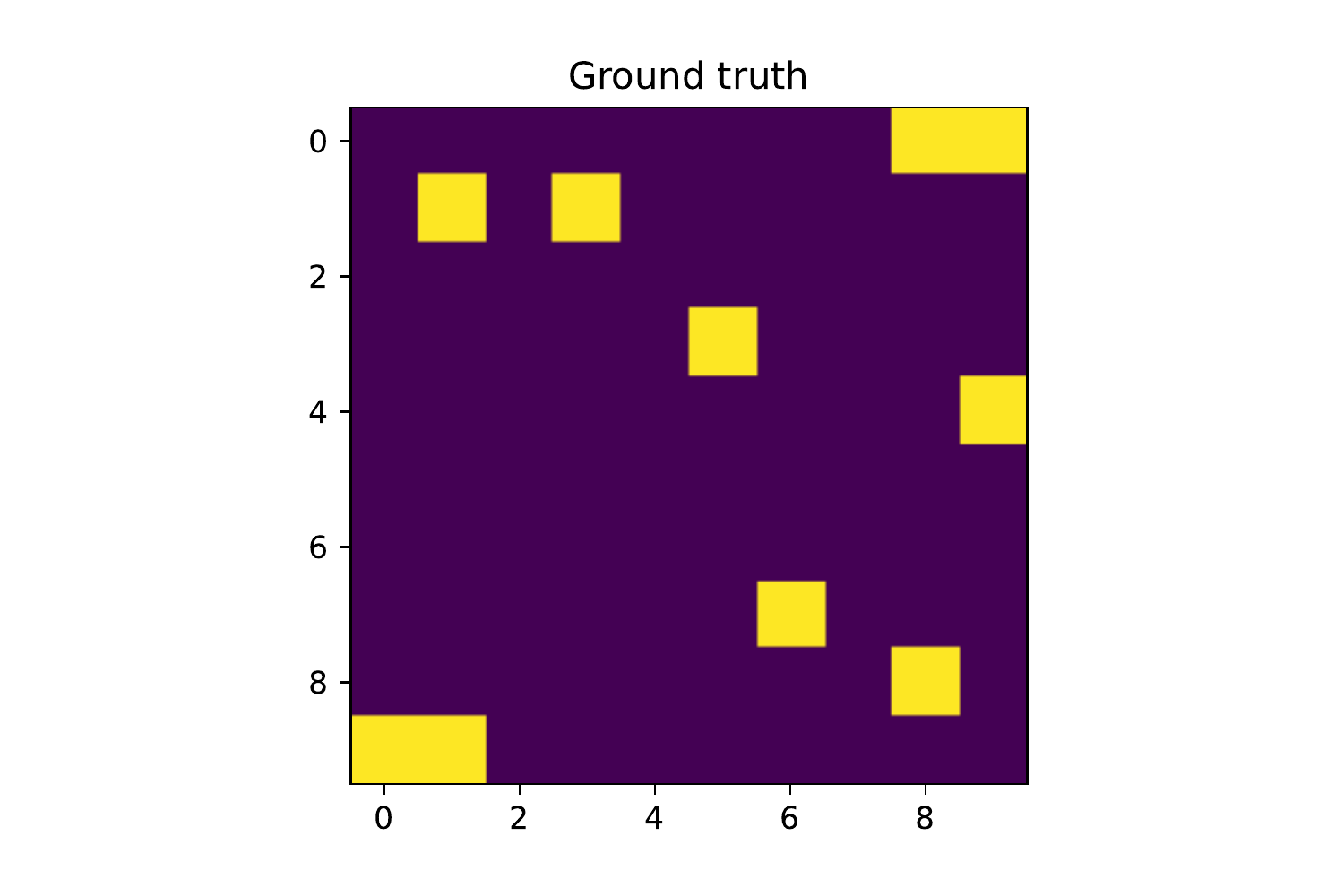}
\caption{\textbf{Simulated ground truth.} This binary mask locates the simulated signal on a 2D $10 \times 10$ grid. Signal locations have been drawn randomly and account for $(1 - \pi_0) \%$ of the image, the rest of the image being null data.}
\label{fig:sim_mask}
\end{figure}

\begin{figure}[H]
\centering
\includegraphics[scale=0.6]{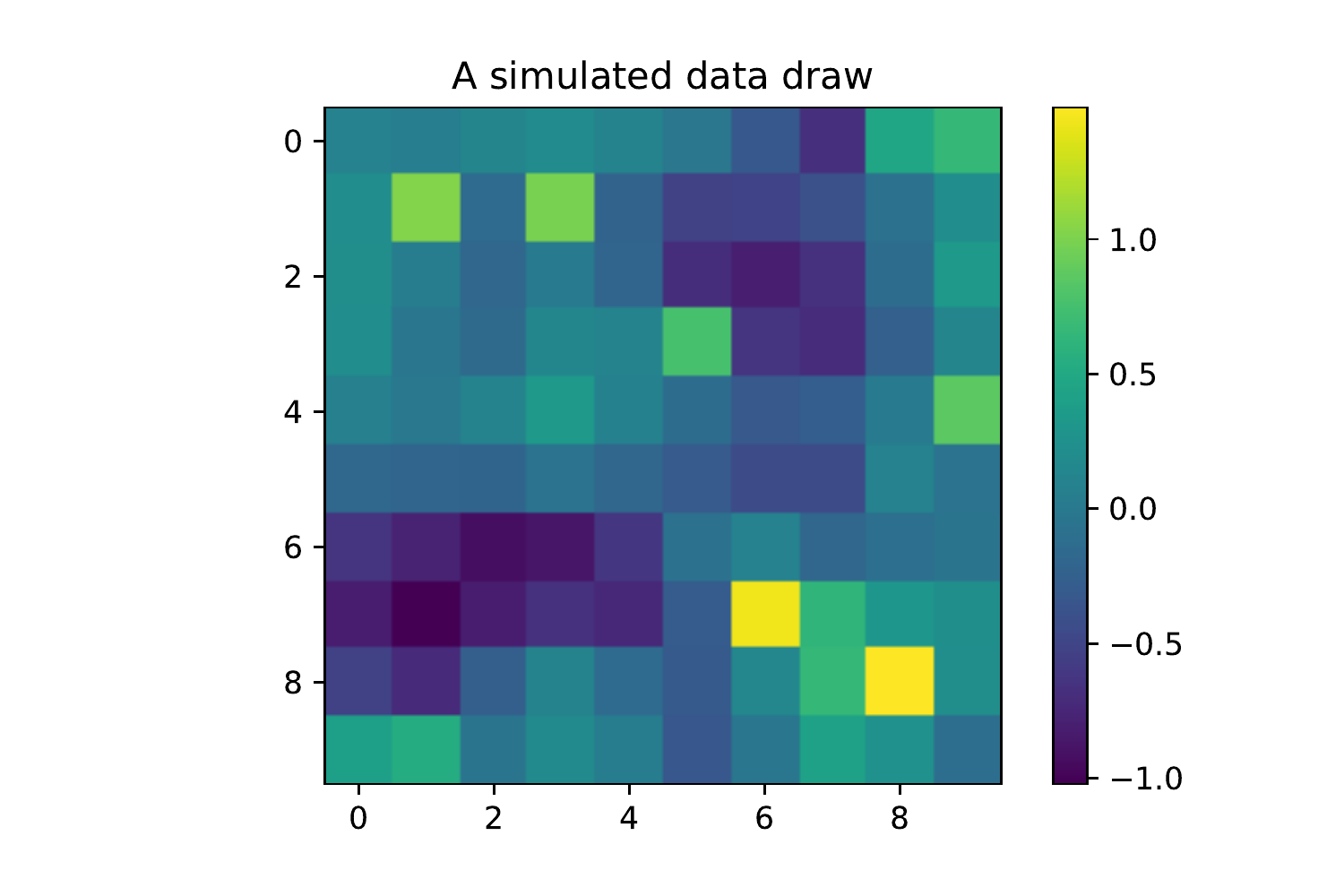}
\caption{\textbf{A simulation draw.} This 2D $10 \times 10$ grid represents a draw of one-sample image comprising signal at locations determined by the binary mask shown in Figure \ref{fig:sim_mask}. This is a typical example of input data in experiments on simulated data; the goal is then to recover the binary mask using inference methods such as Notip, ARI or calibrated Simes.}
\label{fig:sim_draw}
\end{figure}

\add{Here is an example of simulated data computed in 2D for clarity. We use 3D images in the experiments to mimick fMRI data. Here, we use a $10 \times 10$ 2D grid and generate the ground truth, a binary mask that defines the signal. Then, we generate $n_{infer}$ null images and $n_{infer}$ images that comprise signal. Substracting these two sets of images results in a list of $n_{infer}$ one-sample images, as in fMRI experiments. In Figure \ref{fig:sim_mask} an example of simulated ground truth is displayed, while Figure \ref{fig:sim_draw} shows an example of simulated one-sample image. Figure \ref{fig:sim_draw} is a noisy version of the ground truth shown in Figure \ref{fig:sim_mask}.}

\begin{table*}[]
\centering
\begin{tabular}{l|l|l|l|l|l|l|l|l}
&&&&&& \multicolumn{3}{c}{True Discovery Proportion}  \\
Cluster ID & X     & Y     & Z     & Peak Stat & Cluster Size (mm3) & ARI & Calibrated Simes & Notip \\ \hline
1          & -33.0 & -94.0 & -17.0 & 5.63      & 7695               & 0.17      & 0.24                   & 0.26          \\ 
1a         & -45.0 & -79.0 & -26.0 & 4.56      &                    &           &                        &               \\ 
1b         & -48.0 & -61.0 & -26.0 & 4.13      &                    &           &                        &               \\ 
1c         & -51.0 & -64.0 & -35.0 & 4.08      &                    &           &                        &               \\ \hline
2          & 66.0  & 2.0   & 16.0  & 5.47      & 14877              & 0.20      & 0.33                   & 0.45          \\ 
2a         & 69.0  & -22.0 & 10.0  & 4.67      &                    &           &                        &               \\ 
2b         & 69.0  & -10.0 & 13.0  & 4.59      &                    &           &                        &               \\
2c         & 69.0  & -28.0 & 13.0  & 4.43      &                    &           &                        &               \\ \hline
3          & -12.0 & -82.0 & -8.0  & 5.40      & 14445              & 0.27      & 0.38                   & 0.50          \\ 
3a         & 30.0  & -73.0 & -8.0  & 4.96      &                    &           &                        &               \\ 
3b         & -24.0 & -61.0 & -11.0 & 4.91      &                    &           &                        &               \\ 
3c         & 30.0  & -46.0 & -11.0 & 4.64      &                    &           &                        &               \\ \hline
4          & -6.0  & 11.0  & 52.0  & 5.30      & 5238               & 0.14      & 0.25                   & 0.29          \\ 
4a         & 6.0   & 8.0   & 55.0  & 4.19      &                    &           &                        &               \\ \hline
5          & 45.0  & 14.0  & 25.0  & 5.27      & 4563               & 0.24      & 0.30                   & 0.30          \\ 
5a         & 48.0  & 29.0  & 13.0  & 3.36      &                    &           &                        &               \\ \hline
6          & 12.0  & -43.0 & -26.0 & 5.08      & 12555              & 0.05      & 0.17                   & 0.35          \\ 
6a         & 0.0   & -64.0 & -14.0 & 4.43      &                    &           &                        &               \\ 
6b         & 3.0   & -55.0 & -11.0 & 4.26      &                    &           &                        &               \\ 
6c         & 3.0   & -16.0 & -32.0 & 4.23      &                    &           &                        &               \\ \hline
7          & 39.0  & -73.0 & 4.0   & 5.00      & 6075               & 0.04      & 0.09                   & 0.17          \\ 
7a         & 39.0  & -64.0 & 16.0  & 4.44      &                    &           &                        &               \\ 
7b         & 30.0  & -82.0 & 10.0  & 4.42      &                    &           &                        &               \\ 
7c         & 27.0  & -67.0 & 34.0  & 3.63      &                    &           &                        &               \\ \hline
8          & -63.0 & -34.0 & 16.0  & 4.95      & 25812              & 0.30      & 0.48                   & 0.66          \\ 
8a         & -63.0 & -10.0 & 13.0  & 4.90      &                    &           &                        &               \\ 
8b         & -27.0 & -19.0 & 4.0   & 4.85      &                    &           &                        &               \\ 
8c         & -57.0 & -19.0 & 7.0   & 4.68      &                    &           &                        &               \\ \hline
9          & 36.0  & -94.0 & -8.0  & 4.75      & 6507               & 0.08      & 0.15                   & 0.17          \\ 
9a         & 48.0  & -70.0 & -32.0 & 3.96      &                    &           &                        &               \\ 
9b         & 45.0  & -70.0 & -23.0 & 3.92      &                    &           &                        &               \\ 
9c         & 33.0  & -82.0 & -29.0 & 3.77      &                    &           &                        &               \\
\end{tabular}

    \caption{\textbf{Cluster localization (z > 3), size, peak statistic and \del{estimated} TDP \add{lower bound} at risk level $\alpha = 5\%$} using the three possible templates (ARI, Calibrated Simes and Notip) on contrast pair 'look negative cue vs look negative rating'. Cluster subpeaks are also reported when relevant. This table can be generated using script \url{https://github.com/alexblnn/Notip/blob/master/scripts/table_2.py}.}
    \label{tab:clusters}
\end{table*}

\begin{table*}[]
\centering
\begin{tabular}{l|l|l|l|l|l|l|l|l}
&&&&&& \multicolumn{3}{c}{True Discovery Proportion}  \\
Cluster ID & X     & Y     & Z     & Peak Stat & Cluster Size (mm3) & ARI & Calibrated Simes & Notip \\ \hline
1          & 66.0  & 2.0   & 16.0  & 5.47      & 14877              & 0.20      & 0.33                   & 0.45          \\ 
1a         & 69.0  & -22.0 & 10.0  & 4.67      &                    &           &                        &               \\ 
1b         & 69.0  & -10.0 & 13.0  & 4.59      &                    &           &                        &               \\
1c         & 69.0  & -28.0 & 13.0  & 4.43      &                    &           &                        &               \\ \hline
2          & -12.0 & -82.0 & -8.0  & 5.40      & 14445              & 0.27      & 0.38                   & 0.50          \\ 
2a         & 30.0  & -73.0 & -8.0  & 4.96      &                    &           &                        &               \\ 
2b         & -24.0 & -61.0 & -11.0 & 4.91      &                    &           &                        &               \\ 
2c         & 30.0  & -46.0 & -11.0 & 4.64      &                    &           &                        &               \\ \hline
3          & -63.0 & -34.0 & 16.0  & 4.95      & 25812              & 0.30      & 0.48                   & 0.66          \\ 
3a         & -63.0 & -10.0 & 13.0  & 4.90      &                    &           &                        &               \\ 
3b         & -27.0 & -19.0 & 4.0   & 4.85      &                    &           &                        &               \\ 
3c         & -57.0 & -19.0 & 7.0   & 4.68      &                    &           &                        &               \\
\end{tabular}

    \caption{\add{\textbf{Cluster localization (z > 3), size, peak statistic and TDP lower bound at risk level $\alpha = 5\%$} using the three possible templates (ARI, Calibrated Simes and Notip) on contrast pair 'look negative cue vs look negative rating'. Cluster subpeaks are also reported when relevant. Notice that we retained 3 clusters (originally of indices 2, 3 and 8 of Table \ref{tdpclusters}).}}
    \label{tab:clusters0}
\end{table*}

\begin{table*}[]
\centering
\begin{tabular}{l|l|l|l|l|l|l|l|l}
&&&&&& \multicolumn{3}{c}{True Discovery Proportion}  \\
Cluster ID & X     & Y     & Z     & Peak Stat & Cluster Size (mm3) & ARI & Calibrated Simes & Notip \\ \hline
1   & 66.0  & 2.0   & 16.0  & 5.47 & 28593 & 0.13 & 0.18 & 0.29 \\
1a  & 69.0  & -22.0 & 10.0  & 4.67 &       &      &      &      \\
1b  & 69.0  & -10.0 & 13.0  & 4.59 &       &      &      &      \\
1c  & 69.0  & -28.0 & 13.0  & 4.43 &       &      &      &      \\ \hline
2   & -12.0 & -82.0 & -8.0  & 5.40 & 23355 & 0.19 & 0.23 & 0.35 \\
2a  & 30.0  & -73.0 & -8.0  & 4.96 &       &      &      &      \\
2b  & -24.0 & -61.0 & -11.0 & 4.91 &       &      &      &      \\
2c  & 30.0  & -46.0 & -11.0 & 4.64 &       &      &      &      \\ \hline
3   & -63.0 & -34.0 & 16.0  & 4.95 & 43092 & 0.19 & 0.25 & 0.42 \\
3a  & -63.0 & -10.0 & 13.0  & 4.90 &       &      &      &      \\
3b  & -27.0 & -19.0 & 4.0   & 4.85 &       &      &      &      \\
3c  & -57.0 & -19.0 & 7.0   & 4.68 &       &      &      &      \\
\end{tabular}

    \caption{\add{\textbf{Cluster localization (z > 2.5), size, peak statistic and TDP lower bound at risk level $\alpha = 5\%$} using the three possible templates (ARI, Calibrated Simes and Notip) on contrast pair 'look negative cue vs look negative rating'. Cluster subpeaks are also reported when relevant.}}
    \label{tab:clusters1}
\end{table*}

\begin{table*}[]
\centering
\begin{tabular}{l|l|l|l|l|l|l|l|l}
&&&&&& \multicolumn{3}{c}{True Discovery Proportion}  \\
Cluster ID & X     & Y     & Z     & Peak Stat & Cluster Size (mm3) & ARI & Calibrated Simes & Notip \\ \hline
1  & 66.0  & 2.0   & 16.0  & 5.47 & 7425 & 0.38 & 0.48 & 0.69 \\
1a & 69.0  & -22.0 & 10.0  & 4.67 &      &      &      &      \\
1b & 69.0  & -10.0 & 13.0  & 4.59 &      &      &      &      \\
1c & 69.0  & -28.0 & 13.0  & 4.43 &      &      &      &      \\
\hline
2  & -12.0 & -82.0 & -8.0  & 5.40 & 8397 & 0.46 & 0.53 & 0.73 \\
2a & 30.0  & -73.0 & -8.0  & 4.96 &      &      &      &      \\
2b & -24.0 & -61.0 & -11.0 & 4.91 &      &      &      &      \\
2c & 30.0  & -46.0 & -11.0 & 4.64 &      &      &      &      \\
\hline
3  & -63.0 & -34.0 & 16.0  & 4.95 & 9585 & 0.46 & 0.55 & 0.76 \\
3a & -63.0 & -10.0 & 13.0  & 4.90 &      &      &      &      \\
3b & -57.0 & -19.0 & 7.0   & 4.68 &      &      &      &      \\
3c & -60.0 & -49.0 & 25.0  & 4.59 &      &      &      & 
\end{tabular}

    \caption{\add{\textbf{Cluster localization (z > 3.5), size, peak statistic and TDP lower bound at risk level $\alpha = 5\%$} using the three possible templates (ARI, Calibrated Simes and Notip) on contrast pair 'look negative cue vs look negative rating'. Cluster subpeaks are also reported when relevant.}}
    \label{tab:clusters2}
\end{table*}

\begin{table*}
\centering

\begin{tabular}{l|l|l|l}
Study               & Contrast 1                                          & Contrast 2     & $n_{subjects}$                                    \\ \hline
HCP                       & shapes vs baseline                  &   faces vs baseline     & 66              \\
HCP                       & right hand vs baseline             &  right foot vs baseline  & 67           \\
HCP                       &   right foot vs baseline             &   left foot vs baseline  & 66            \\
HCP                       &   left hand vs baseline              &   right foot vs baseline & 67            \\
HCP                       &   left hand vs baseline              &   left foot vs baseline   & 66           \\
HCP                       &   tool vs baseline                    &   face vs baseline      & 68              \\
HCP                       &   face vs baseline                    &   body vs baseline       & 68             \\
HCP                       &   tool vs baseline                    &   body vs baseline  & 68                  \\
HCP                       &   body vs baseline                    &   place vs baseline   & 68                \\
amalric2012mathematicians &   equation vs baseline                &   number vs baseline & 29                 \\
amalric2012mathematicians &   house vs baseline                   &   word vs baseline   & 37                 \\
amalric2012mathematicians &   house vs baseline                   &   body vs baseline   & 27                 \\
amalric2012mathematicians &   equation vs baseline                &   word vs baseline    & 29                \\
amalric2012mathematicians &   visual calculation vs baseline     &   auditory sentences vs  baseline & 27    \\
amalric2012mathematicians &   auditory right motor vs baseline  &   visual calculation vs  baseline & 25    \\
cauvet2009muslang         &  c16 music vs baseline              &    c02 music vs baseline & 35             \\
cauvet2009muslang         &    c16 language vs baseline           &   c01  language vs baseline & 35           \\
cauvet2009muslang         &    c02 language vs baseline           &   c16 language vs baseline & 35           \\
cauvet2009muslang         &    c04 language vs baseline           &   c16 language vs baseline & 35           \\
amalric2012mathematicians &   face vs baseline                    &   scramble vs baseline  & 85              \\
ds107                     &   scramble vs baseline                &   objects vs baseline & 44                \\
ds107                     &   consonant vs baseline               &   scramble vs baseline  & 47              \\
ds107                     &   consonant vs baseline               &   objects vs baseline   & 44              \\
ds108                     &   reapp negative rating vs baseline &   reapp negative cue vs baseline & 32   \\
ds108                     &   look negative stim vs baseline    &   look negative rating vs baseline  & 34\\
ds108                     &   reapp negative stim vs baseline   &   reapp negative rating vs baseline & 34\\
ds109                     &   false photo story vs baseline     &   false photo question vs baseline & 36 \\
ds109                     &   false belief story vs baseline    &   false photo story vs baseline  & 36   \\
ds109                     &   false belief question vs baseline &   false photo question vs baseline & 36 \\
ds109                     &   false belief story vs baseline    &   false belief question vs baseline & 36\\
ds109                     &   false belief question vs baseline &   false photo story vs baseline  & 36   \\
pinel2007fast             &   visual right motor vs baseline    &   vertical checkerboard vs baseline  & 113\\
pinel2007fast             &   auditory right motor vs baseline  &   visual right motor vs baseline   & 121 \\
ds107                     &   scramble vs baseline                &   face vs baseline    & 85                \\
amalric2012mathematicians &   house vs baseline                   &   scramble vs baseline & 85               \\
ds107                     &   words vs baseline                   &   face vs baseline   & 100                
\end{tabular}
    \caption{\textbf{36 pairs of fMRI contrasts used for experiments.} These contrasts images have been downloaded from Neurovault 1952 collection.}
    \label{tab:contrasts}
\end{table*}

\end{document}